\documentstyle[prd,aps,preprint,psfig]{revtex}

\newcommand{\ev}{ {\rm eV} }

\newcommand{\s}{ {\rm s} }

\newcommand{\anti}[1]{{\overline{#1}}}

\begin{document}

%@twocolumnfalse\endcsname 

%%
\tighten

\preprint{
\noindent
\begin{minipage}[t]{3in}
\begin{flushright}
RESCEU-02-4\\
UTAP-02-411\\
YITP-02-16\\
astro-ph/0203481 \\
\end{flushright}
\end{minipage}
}

\draft
\title{%%
Gamma-Ray Burst Neutrino Background and Star Formation History in the Universe
}%%

\author{Shigehiro Nagataki$^1$, Kazunori Kohri$^2$, Shin'ichiro Ando$^{1}$,
and Katsuhiko Sato$^{1,3}$}

\address{%%
$^1$Department of Physics, School of Science, University
of Tokyo, 7-3-1 Hongo, Bunkyoku, Tokyo 113-0033, Japan
{\tt nagataki@utap.phys.s.u-tokyo.ac.jp}
}%%

\address{%%
$^2$Yukawa Institute for Theoretical Physics, Kyoto University, 
Kyoto, 606-8502, Japan
%%{\tt kohri@yukawa.kyoto-u.ac.jp}
}%%

\address{%%
$^3$Research Center for the Early Universe, University
of Tokyo, 7-3-1 Hongo, Bunkyoku, Tokyo 113-0033, Japan
}%%

%\date{\today}

\maketitle

\begin{abstract}
    We estimate the flux of the GRB neutrino background and compute
    the event rate at SK and TITAND in the collapsar model, assuming
    that GRB formation rate is proportional to the star formation
    rate. We find that the flux and the event rate depend sensitively
    on the mass-accretion rate.  Although the detection of signals
    from GRBs seems to be difficult by SK, we find that we can detect
    them by TITAND (optimistically $\sim$~20 events per year in the
    energy range $E_{\anti{\nu}_e} \ge 50$MeV).  Therefore, we show
    that we will obtain the informations on the mass-accretion rate of
    collapsar by observing the GRB neutrino background, which, in
    turn, should give us informations on the total explosion energy of
    GRBs.
\end{abstract}

%\hspace{1.5cm} UTAP-***, hep-ph/yymmxxx}%%

%98.62.Mw Infall, accretion, and accretion disks
%97.60.-s Late stages of stellar evolution
%98.70.Rz Gamma-ray sources; gamma-ray bursts
%95.85.Ry Neutrino, muon, pion, and other elementary particles; cosmic rays

%]

%%%%%%%%%%%%%%%%%%%%%%%%%%%%%%%%%%%%%%%%%%%%%%%%%%%%%%%%%%%%%%%%%%%%%%
\section{Introduction}
%%%%%%%%%%%%%%%%%%%%%%%%%%%%%%%%%%%%%%%%%%%%%%%%%%%%%%%%%%%%%%%%%%%%%%
%\vspace{2cm}

Gamma-Ray Burst (GRB) is one of the most mysterious phenomena in the
universe. Although there was a great revolution in our understanding
on GRBs in 1997 that some of them were proved to be of extragalactic
origins~\cite{boella97}, there are still a lot of questions and
puzzles left on the phenomena.  That is why GRBs have been the
fascinating targets for astronomers.  In fact, there are a lot of
interesting and valuable discussions under debate. For example, we
have not known what is the origin of GRB.  Although this problem has
never been settled and still controversial, there are some reports
that suggest the physical connections between GRBs and supernovae
(SNe)~\cite{galama98}. If the SN/GRB connection is correct, GRBs will
be born from the death of massive stars. Then the GRB event rate
should trace the star formation history in the
universe~\cite{porciani01}. 

The central engine of a GRB is an important but difficult problem to
be solved. Although a fireball model~\cite{meszaros97} is a promising
and attractive one, we do not know how the initial condition is. We
merely gave it artificially.  In the fireball model, since the ratio
of baryon number to photon number is required to be extremely small as
an initial condition, only a peculiar environment could meet such a
condition.  As long as we believe the GRB/SN connection, however, a
collapsar (failed supernova) model~\cite{woosley93} will be one of the
convincing candidates as the central engine of a GRB. In the model,
much energy is generated by the neutrino heating from an accretion
disk surrounding a black hole which is made by the core collapse of a
massive star.

On the other hand, total explosion energy of a GRB is also one of the
most important problems in the physics of GRBs. Observationally it is
well known that there is a large dispersion of total energy of
gamma-rays emitted from GRBs~\cite{bloom01}. Although the dispersion
may become small when we consider the beaming effects~\cite{frail01},
after all we do not know the total explosion energy of the system that
generates a GRB only by the gamma-ray observations.\footnote{ %%
Here, the total explosion energy means the summation of energies of
photons, leptons and baryons emitted from an origin of a GRB except
for neutrinos.
} %%
In the theoretical side, the total explosion energy in the collapsar
model is also not estimated clearly, and only a rough estimate is
reported as a result of the numerical computations~\cite{macfadyen99}.
Therefore, to know the total explosion energy, we require the
multi-dimension numerical hydrodynamical computations including the
microphysics of neutrino heating in future, which is, of course, a
laborious task to be performed.

Are there observations which give us new clues to know how a GRB is
produced?  In this study, we point out that the GRB neutrino
background may be detected in a large water Cherenkov neutrino
detector such as Super-Kamiokande (SK) and TITAND which is proposed as
a next-generation plan of the multi-megaton water Cherenkov
detector~\cite{suzuki:2001}. We also stress that this signal should
give us valuable informations on GRBs, such as the total explosion
energy and/or event rate of GRBs.  The reason is as follows.  In this
study, we assume that the origins of GRBs are massive stars.  Thus,
the event rate of GRBs should trace the star formation history in the
universe. Since we adopt the collapsar model here, we can assume that
the central engine of a GRB is the neutrino heating from an accretion
disk surrounding the central black hole.  We show that the total
number of neutrinos emitted from a GRB is very sensitive to the
mass-accretion rate $\dot{M}$ which is difficult to estimate only from
the existing theoretical models.  However, we can obtain the
informations of the mass-accretion rate by observing the GRB neutrino
background.  We also derive an approximate relationship between the
total explosion energy and the mass-accretion rate of a GRB.  Namely,
although we cannot determine the total explosion energy only from the
gamma-ray observations, we can obtain informations of it
by observing the GRB neutrino background. That is an important clue to
understand GRBs.  In this study we find that the GRB neutrino
background may be detected within $\sim$ 10 yrs by TITAND as long as
the average mass-accretion rate among GRBs is $\dot{M} \gtrsim$ a few
$M_{\odot}$ s$^{-1}$, and the probability that one collapsar generates
a GRB is $f$ $\sim$ 0.5~--~1.0.
%Of course, a
%larger water Cherenkov detector than TITAND will detect the signals
%from GRBs more immediately even if the GRBs obey the conservative
%parameters.  

Here we outline the organization of this paper. In
section~\ref{formula}, we explain the formulations in this study.
Results are shown in section~\ref{results}. Discussions are given in
section~\ref{discussion}. Summary and conclusion are presented in
section~\ref{summary}.

%%%%%%%%%%%%%%%%%%%%%%%%%%%%%%%%%%%%%%%%%%%%%%%%%%%
\section{Formulation}\label{formula}
%%%%%%%%%%%%%%%%%%%%%%%%%%%%%%%%%%%%%%%%%%%%%%%%%%%
%%%%%%%%%%%%%%%%%%%%%%%%%%%%%%%%%%%%%%%%%%%%%%%%%%%
\subsection{GRB formation history}
%%%%%%%%%%%%%%%%%%%%%%%%%%%%%%%%%%%%%%%%%%%%%%%%%%%
First we assume that the rate of GRBs traces the global star formation
history of the universe, $R_{\rm GRB} (z) \propto R_{\rm SFR}$.
Although a number of workers have modeled the expected evolution of
the cosmic SFR with redshift, there are some uncertainties (in
particular, at high redshift $z \ge 1 $), and further observations are
required to determine the cosmic SFR. In this study, we use the three
different parameterizations of the global star formation rate per
comoving volume in an Einstein-de Sitter universe to take into account
the uncertainty of the cosmic SFR which is derived by
Ref.~\cite{porciani01}. They are
\begin{eqnarray}
    R_{\rm SF1} (z) = 0.3 h_{65} \frac{\exp (3.4 z)}{ \exp (3.8 z) + 45}
          \; M_{\odot} \rm yr^{-1} Mpc^{-3},
    \label{eq:1}
\end{eqnarray}
\begin{eqnarray}
    R_{\rm SF2} (z) = 0.15 h_{65} \frac{\exp (3.4 z)}{ \exp (3.4 z) + 22}
          \; M_{\odot} \rm yr^{-1} Mpc^{-3},
    \label{eq:2}
\end{eqnarray}
and
\begin{eqnarray}
    R_{\rm SF3} (z) = 0.2 h_{65} \frac{\exp (3.05 z  - 0.4)}{ \exp (2.93 z) 
                     + 15}
          \; M_{\odot} \rm yr^{-1} Mpc^{-3},
    \label{eq:3}
\end{eqnarray}
where $h_{65} = H_0 / \rm 65 \; km \;s^{-1} \; Mpc^{-1}$.

Assuming that massive stars with masses larger than $\sim 35
M_{\odot}$ explode as GRBs~\cite{macfadyen99}~\cite{woosley95},
GRB-rate density $R_{\rm GRB}$ can then be estimated by multiplying
the selected SFR by the coefficient
\begin{eqnarray}
    f \times \frac{\int^{125}_{35} dm \; \phi (m)}{ \int^{125}_{0.4}
    dm \; m \phi (m) } = 1.5 \times 10^{-4} \left( \frac{f}{0.1}
    \right) M_{\odot}^{-1},
    \label{eq:4}
\end{eqnarray}
where $\phi (m)$ is the initial mass function (IMF) and $m$ is the
stellar mass in solar units. In this study, the Salpeter's IMF ($\phi
(m) \propto m^{-2.35}$) is adopted throughout.  
Here we assumed that one tenth
of the collapsars whose mass range is in
(35 -- 125)$M_{\odot}$ are accompanied with GRBs ($\equiv f$).
The validity of this
assumption and the detailed discussions are presented in
section~\ref{discussion}.
We label these models
as GRBR1, GRBR2, and GRBR3, respectively.  For comparison, we use the
SN-rate density $R_{\rm SN}$ by multiplying the selected SFR by the
coefficient
\begin{eqnarray}
        \frac{\int^{35}_{8} dm \; \phi (m)}{ \int^{125}_{0.4} dm \; 
        m \phi (m)  }
        = 1.2 \times 10^{-2} M_{\odot}^{-1}.
    \label{eq:4.2}
\end{eqnarray}
As for the energy spectrum of $\anti{\nu}_e$ of SNe, the Fermi-Dirac
distribution with zero chemical potential is used (see
Figs~\ref{flux1}--~\ref{flux3}), which have the same luminosity with
the numerical model~\cite{ando02}.

\begin{figure}[htbp]
\centerline{\psfig{figure=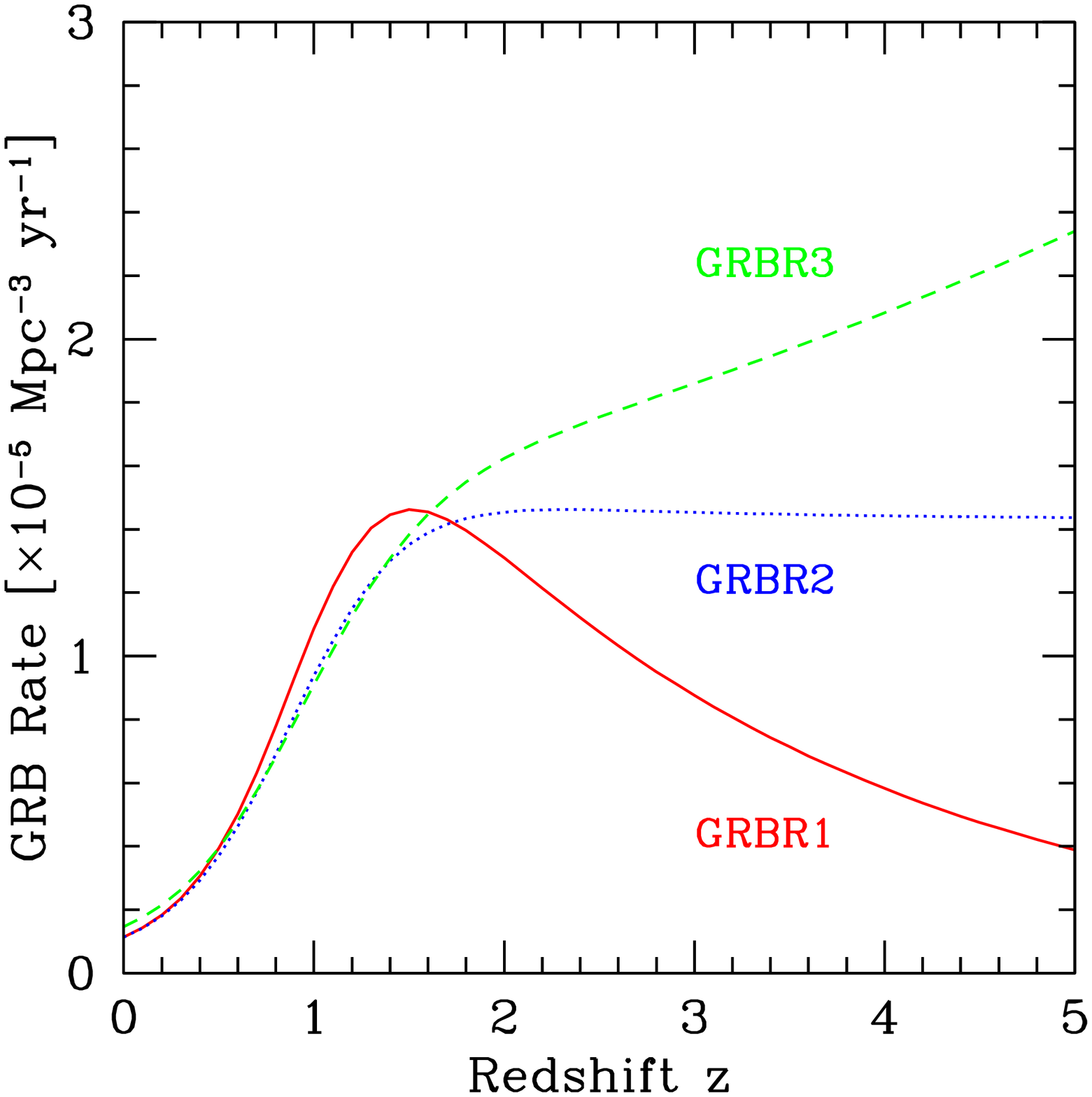,width=15cm}}
\caption{ %%
The cosmic GRB formation history. The solid, dotted and dashed curves
show the rate of GRB formation per unit comoving volume as function of
redshift for the models GRBR1, GRBR2, and GRBR3,
respectively. $\Lambda$-dominated cosmology ($\Omega_M$=0.3,
$\Omega_{\Lambda}$=0.7) is adopted. The Hubble constant is taken to be
70 km s$^{-1}$ Mpc$^{-1}$. The probability that one collapsar
generates a GRB ($f$) is set to be 0.1.
} %%
\label{fig1}
\end{figure}

In the flat universe, i.e., when the density parameter is $\Omega =
\Omega_M + \Omega_{\Lambda} = 1$ with the non-relativistic matter
density ($\Omega_M \sim 0.3 $) and the $\Lambda$ term
($\Omega_{\Lambda} \sim 0.7$), the star formation rate is given
by~\cite{porciani01}
\begin{eqnarray}
        R_{\rm SF} (z; \Omega_M. \Omega_{\Lambda}, h_{65})
         = h_{65}
%        &R&_{\rm SF} (z; \Omega_M. \Omega_{\Lambda}, h_{65}) \\ \nonumber 
%         &=& h_{65}
        \frac{\sqrt{( (1+ \Omega_M z)  ) (1+z)^2 - \Omega_{\Lambda}
        (2z + z^2)}}{(1+z)^{3/2}} R_{\rm SF} (z; 1, 0 ,1 ),
    \label{eq:5}
\end{eqnarray} 
where $z$ is redshift. In Fig.~\ref{fig1}, we show the GRB rates as a
function of the redshift.

%%%%%%%%%%%%%%%%%%%%%%%%%%%%%%%%%%%%%%%%%%%%%%%%%%%
\subsection{Accretion disk model}
%%%%%%%%%%%%%%%%%%%%%%%%%%%%%%%%%%%%%%%%%%%%%%%%%%%
%%%%%%%%%%%%%%%%%%%%%%%%%%%%%%%%%%%%%%%%%%%%%%%%%%%
\subsubsection{Energy Spectrum}\label{enespe}
%%%%%%%%%%%%%%%%%%%%%%%%%%%%%%%%%%%%%%%%%%%%%%%%%%%
In the collapsar model, neutrinos are mainly generated at the
inner-most region of an accretion disk which is formed around the
central black hole. Thus we estimate the energy spectrum of neutrinos
using an analytical shape of the accretion disk. Popham et al. (1999)
proposed an accretion-disk model including the effects of general
relativity as a collapsar model~\cite{popham99}. The model in
Ref.~\cite{popham99} and the numerically computed collapsar
model~\cite{macfadyen99} are fitted well by the analytical shape of
the accretion disk
in Ref~\cite{popham99,fujimoto:2001}~\footnote{ %%
On the other hand, in the merger models, e.g., for the merger of NS-NS
binaries, the analytical equilibrium solutions of the neutrino-cooled
accretion disk and the properties of the neutrino emission are
discussed in Ref.~\cite{ref:kohri02}.
}. %%
We use it in this study.  Then, the density [g cm$^{-3}$], the
temperature [MeV] and the disk thickness [cm] are fitted as follows
(see Ref.~\cite{nagataki02} for details).
\begin{eqnarray}
    \label{eq:rho_N}
    \rho = 8.23 \times 10^{8}
%    \rho = 8.23 &\times& 10^{8}
    \left(\frac{M}{3M_{\odot}}\right)^{-1.7} 
    \left(\frac{\dot{M}}{0.1 M_{\odot} {\rm s}^{-1}}\right)^{1.03}
     \frac{1}{ \displaystyle{\left(\frac{r}{r_s}\right)^{1.07}
       \left( 1 + \left( \frac{r}{r_s} \right)  \right)^{0.76}}  } , \\ 
%    \left(\frac{\dot{M}}{0.1 M_{\odot} {\rm s}^{-1}}\right)^{1.03}\\ \nonumber 
%     &\times& \frac{1}{ \left(\frac{r}{r_s}\right)^{1.07}
%       \left( 1 + \left( \frac{r}{r_s} \right)  \right)^{0.76}  } , \\ 
    \label{eq:temperature}
   T = 2.3 \times \left(\frac{M}{3M_{\odot}}\right)^{-0.2}
   \left(\frac{\dot{M}}{0.1 M_{\odot} {\rm s}^{-1}}\right)^{0.108}
%   T = 2.3 &\times& \left(\frac{M}{3M_{\odot}}\right)^{-0.2}
%   \left(\frac{\dot{M}}{0.1 M_{\odot} {\rm s}^{-1}}\right)^{0.108}\\ \nonumber
    \frac{1}{ \displaystyle{\left(\frac{r}{r_s}\right)^{0.425}
       \left( 1 + \left( \frac{r}{r_s} \right)  \right)^{0.21}}  } , \\  
%   &\times& \frac{1}{ \left(\frac{r}{r_s}\right)^{0.425}
%       \left( 1 + \left( \frac{r}{r_s} \right)  \right)^{0.21}  } , \\  
    \label{eq:disk_thicknss}
    H = 5.8 \times 10^{6}  \left(\frac{M}{3M_{\odot}} \right)^{0.9}
\left(\frac{\dot{M}}{0.1 M_{\odot} {\rm s}^{-1}}\right)^{-0.0183}
 \frac{1}{\displaystyle{ \left(\frac{r}{r_s}\right)^{-1.66}
%    H = 5.8 &\times& 10^{6}  \left(\frac{M}{3M_{\odot}} \right)^{0.9}
%\left(\frac{\dot{M}}{0.1 M_{\odot} {\rm s}^{-1}}\right)^{-0.0183}\\ \nonumber
% &\times& \frac{1}{ \left(\frac{r}{r_s}\right)^{-1.66}
          \left( 1 + \left( \frac{r}{r_s} \right)  \right)^{0.3867}} },
\end{eqnarray}
where $M$ is the mass of the central black hole, $r$ is the radial
coordinate, and $r_s$ (= 10$^{7}$cm) is the core radius, respectively.
Note that the Schwarzshild radius is $\simeq 8.862{\rm km} (M/3
M_{\odot})$.  In Ref.~\cite{popham99}, they assumed that the mass of
the central black hole is constant. In this study, however, to
estimate the total energy of neutrinos we consider the effects that
the mass of the central black hole becomes larger with time by
accretions.

\begin{figure}[htbp]
\centerline{\psfig{figure=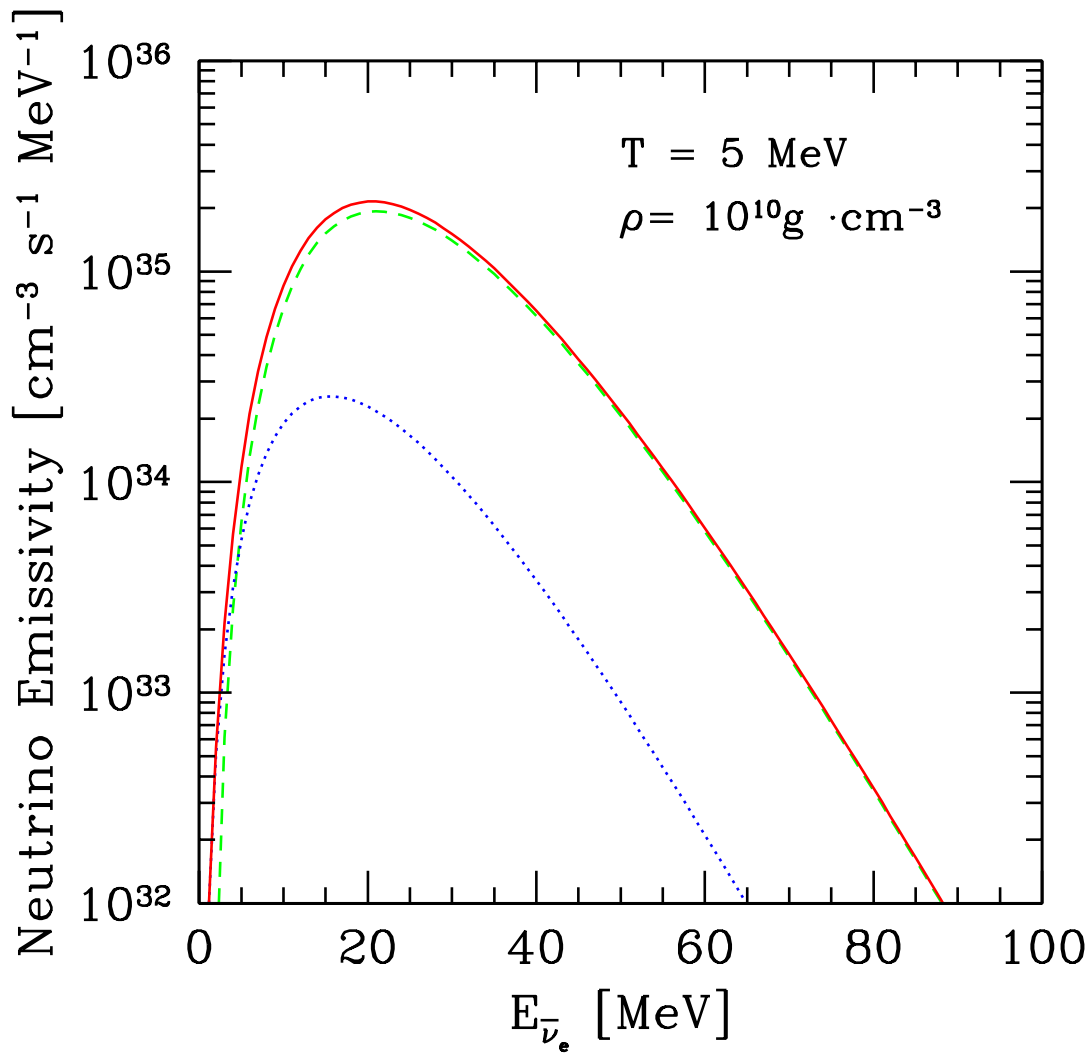,width=15cm}}
\caption{
Emissivity of $\anti{\nu}_e$. Here we adopt $T$ = 5 MeV and
       $\rho=10^{10} {\rm g} \cdot {\rm cm}^{-3}$. Solid line represents the
       total emissivity of $\anti{\nu}_e$.  Dashed (dotted) line is the
       contribution from $n + e^+ \rightarrow p + \anti{\nu}_e$ ($e^+ +
       e^- \rightarrow \nu_e + \anti{\nu}_e$).
}
\label{fig2}
\end{figure}

When we know the density and the temperature in the neutrino emitting
region, we can estimate the emissivity of neutrinos ($\rm cm^{-3}
sec^{-1}MeV^{-1}$) in the accretion disk.  The total emissivity of
$\anti{\nu}_e$ is represented by two terms:
\begin{eqnarray}
    \label{eq:tot_emissivity}
    \frac{d^2n_{\anti{\nu}_e}}{dt dE_{\anti{\nu}_e}} =
    \frac{d^2n^{eN}_{\anti{\nu}_e}}{dt dE_{\anti{\nu}_e}} +
    \frac{d^2n^{e^+e^-}_{\anti{\nu}_e}}{dt dE_{\anti{\nu}_e}},
\end{eqnarray}
where $E_{\anti{\nu}_e}$ is energy of $\anti{\nu}_e$.  For $n + e^+
\rightarrow p + \anti{\nu}_e$, the emissivity of $\anti{\nu}_e$ is
represented by
\begin{eqnarray}
%        \nonumber
    \frac{d^2 n_{\anti{\nu}_e}^{eN}}{dt
    dE_{\anti{\nu}_e}}(E_{\anti{\nu}_e})  = &&
    \frac{G_F^2}{2\pi^3}(1+3g_A^2)n_nE_{\anti{\nu}_e}^2
    \sqrt{\left(E_{\anti{\nu}_e}-Q\right)^2-m_e^2}  
%     \\ &\times& \left(E_{\anti{\nu}_e}-Q\right)
     \left(E_{\anti{\nu}_e}-Q\right)
    \frac1{e^{(E_{\anti{\nu}_e}-Q)/T}+1},
\label{eq:nusp_nuc}
\end{eqnarray}
where $G_F$ is Fermi coupling constant, $g_A$ is the axial vector
coupling constant of the nucleon which is normalized by the
experimental value of neutron lifetime $\tau_n \simeq 886.7$
s~\cite{Groom:in}, $n_n$ is number density of neutron, Q $\simeq$ 1.29
MeV, and $m_e$ is electron mass. For $e^- + e^- \rightarrow \nu_e +
\anti{\nu}_e$, we obtain
\begin{eqnarray}
    \label{eq:nusp_ann}
%        \nonumber
    \frac{d^2 n_{\anti{\nu}_e}^{e^+e^-}}{dt
    dE_{\anti{\nu}_e}}(E_{\anti{\nu}_e})= &&
    \frac{G_F^2}{9\pi^4}(C_V^2+C_A^2)E_{\anti{\nu}_e}^3
    \frac1{e^{E_{\anti{\nu}_e}/T}+1}T^4
%     \\ &\times&    \int^{\infty}_{m_e/T}
     \int^{\infty}_{m_e/T}
    \frac{(\epsilon^2-(m_e/T)^2)^{3/2}}{e^{\epsilon}+1}d \epsilon,
\end{eqnarray}
where $C_V = 1/2+2\sin^2\theta_W$, $C_A$ = 1/2, and $\sin^2\theta_W
\simeq 0.231$ is Weinberg angle~\cite{Groom:in}, and we assume $T \gg
m_e$.  We plot them in Fig.~\ref{fig2} in the case that $T=$ 5MeV and
$\rho=10^{10} {\rm g} \cdot {\rm cm}^{-3}$.

\begin{figure}[htbp]
%\vspace{-0.8cm}
\centerline{\psfig{figure=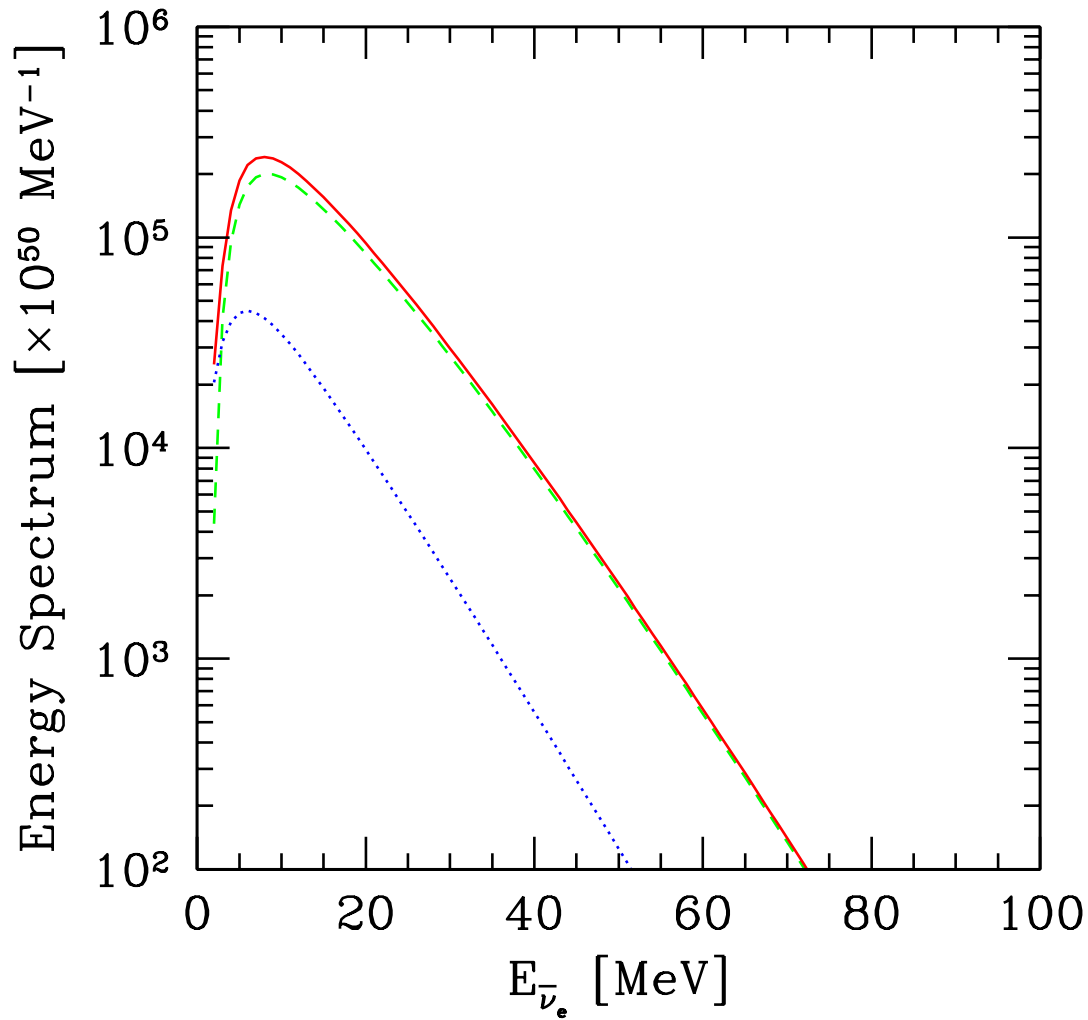,width=15cm}}
%\vspace{-0.5cm}
\caption{
       Energy spectrums of $\anti{\nu}_e$ from a collapsar.  Solid line
       represents the total energy spectrum.  Dashed (dotted) line is
       the contribution from $n + e^+ \rightarrow p + \anti{\nu}_e$
       ($e^+ + e^- \rightarrow \nu_e + \anti{\nu}_e$). The total
       accreting mass, the initial mass, and the mass-accretion rate
       are set to be 30$M_{\odot}$, 3$M_{\odot}$, and 0.1$M_{\odot}$
       $s^{-1}$, respectively.
}
\label{fig3}
\end{figure}

Using Eqs.~(\ref{eq:rho_N}),~(\ref{eq:temperature})
and~(\ref{eq:disk_thicknss}), we obtain the integrated energy
spectrum of $\anti{\nu}_e$,
\begin{eqnarray}
    \label{eq:dndE}
    \frac{d n}{dE_{\anti{\nu}_e}}(E_{\anti{\nu}_e}) = \int_0^{\Delta t}
    dt \int_{T \ge 1 {\rm MeV}}dV_a \frac{d^2 n_{\anti{\nu}_e}}{dt
    dE_{\anti{\nu}_e}}(E_{\anti{\nu}_e}),
\end{eqnarray}
where $\int dV_a$ denotes the integration in the volume of the
emitting region in accretion disk where the temperature is high enough
to produce neutrinos, i.e., $T \geq 1$ MeV, $\int dt$ denotes the
integration of time $t$, and $\Delta t \equiv M_a/\dot{M}$ is the
duration of the accretion ($M_a$ is the mass of the progenitor).
Here we assume that the time evolution of
the mass of the central black hole is given by
\begin{eqnarray}
    \label{eq:mass_evolution}
    M = M_i + \dot{M} t,
\end{eqnarray}
where $M_i$ is the initial value of $M$.
In this study, $M_i$ is set to be 3$M_{\odot}$.
In Fig.~\ref{fig3} we plot
the obtained energy spectrum of $\anti{\nu}_e$ emitted from the
accretion disk in unit energy [${\rm MeV}^{-1}$].
%Note that the high
%energy tail is not dumped at all. That is because the nucleon density
%of the accretion disk is lower than that of a neutron star, and the
%mean free path is longer. This is remarkable feature only for
%collapsars.

\begin{figure}[htbp]
%\vspace{-0.8cm}
\centerline{\psfig{figure=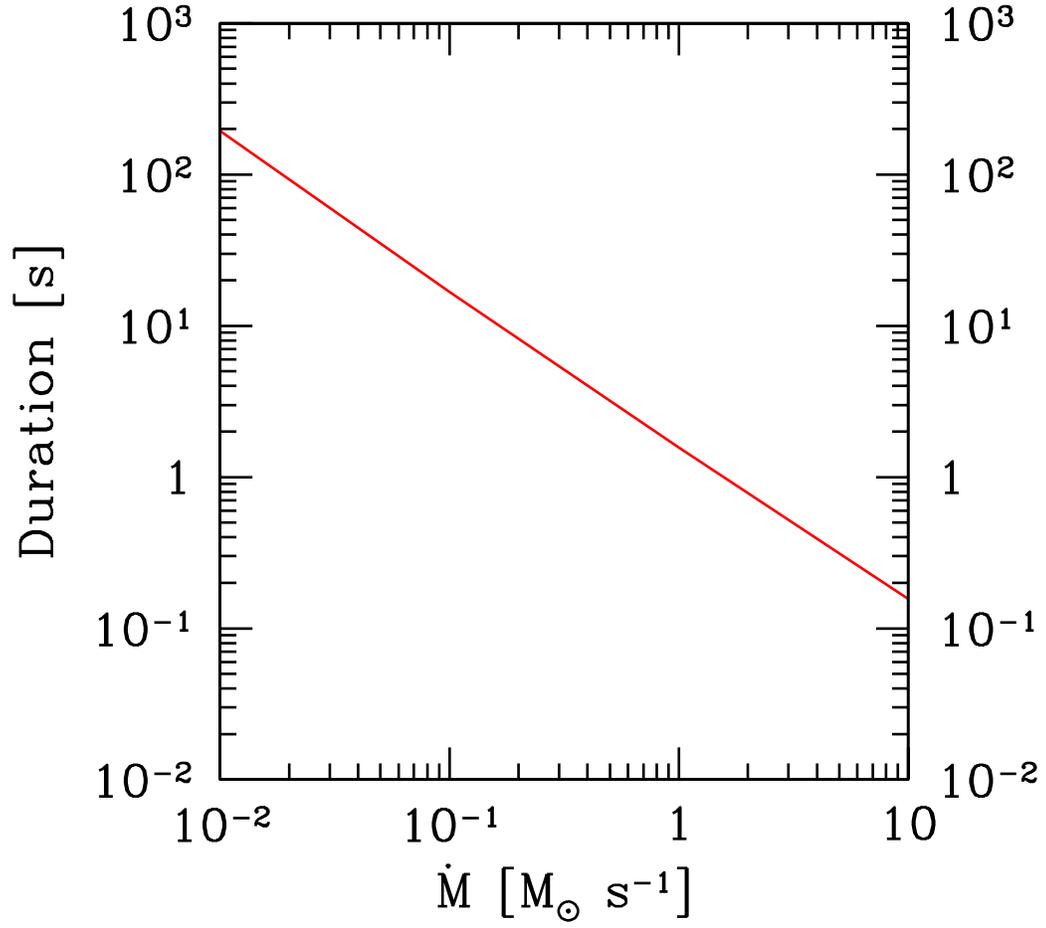,width=15cm}}
%\vspace{-0.5cm}
\caption{
Duration of the neutrino emission from a collapsar as function
of the mass-accretion rate. The total accreting mass and the initial mass
are set to be 30$M_{\odot}$, 3$M_{\odot}$, respectively.
}
\label{duration}
\end{figure}

Here we define the time-dependent luminosity of $\nu_e$ and
$\anti{\nu}_e$ by
\begin{eqnarray}
    \label{eq:L_nu1}
    L_{\nu}(t) = \int_{T \ge 1 {\rm MeV}}dV_a \left(%%
      \int dE_{\anti{\nu}_e}\frac{d^2 n_{\anti{\nu}_e}}{dt
      dE_{\anti{\nu}_e}}(E_{\anti{\nu}_e}) + \int dE_{{\nu_e}}\frac{d^2
      n_{{\nu_e}}}{dt dE_{{\nu_e}}}(E_{{\nu_e}})\right), %%
\end{eqnarray}
where $dn_{\nu_e}/dtdE_{{\nu_e}}$ is the total emissivity of $\nu_e$
with their energy $E_{{\nu_e}}$, and we assume it is approximately
equal to $dn_{\anti{\nu}_e}/dtdE_{{\anti{\nu}_e}}$. In addition, we
define the total energy of $\nu_e$'s and $\anti{\nu}_e$'s emitted from
a collapsar,
\begin{eqnarray}
    \label{eq:q_tot}
    Q_{\rm tot} = \int_0^{\Delta t} L_{\nu}(t)dt.
\end{eqnarray}
We see from Eq.~(\ref{eq:temperature}) that as the mass of the central
black hole grows, the temperature becomes lower, and the flux of
neutrinos decreases. In Fig.~\ref{duration}, we show the duration
($\equiv \Delta t_{\nu,{\rm emit}}$) which is defined as the e-folding
time of the time-dependent luminosity in Eq.~(\ref{eq:L_nu1}). This
almost corresponds to the timescale in which the temperature at the
innermost region of the accretion disk becomes lower than $\sim$ 1
MeV. Namely this will approximately reflect the timescale of a GRB.

\begin{figure}[htbp]
%\vspace{-0.8cm}
\centerline{\psfig{figure=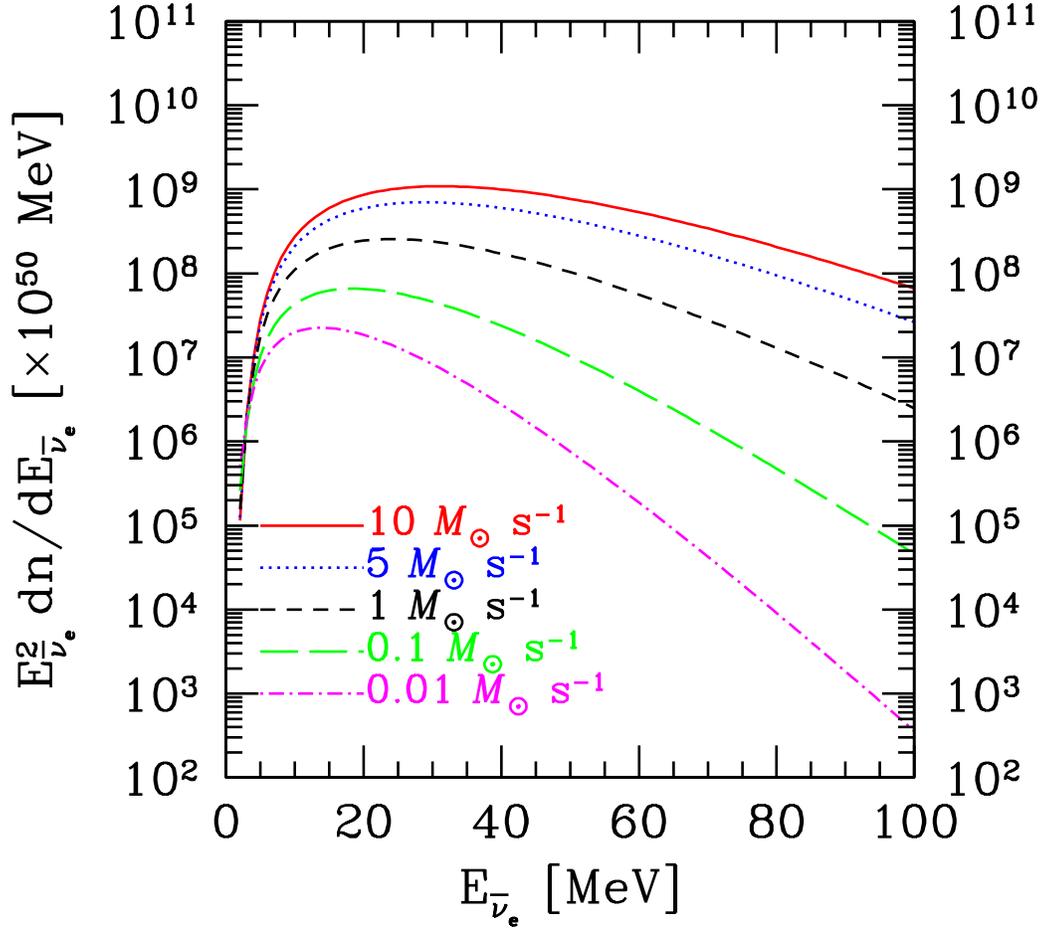,width=15cm}}
%\vspace{-0.5cm}
\caption{ %%
Dependence of the energy spectrum of
anti-electron neutrino on the mass-accretion rate ($\dot{M}$).
Solid line, dotted line, short-dashed line, long-dashed line, and
dot-short-dashed line
correspond to $\dot{M}$ = 10$M_{\odot}$ s$^{-1}$, 5$M_{\odot}$ s$^{-1}$,
1$M_{\odot}$ s$^{-1}$, 0.1$M_{\odot}$ s$^{-1}$ and 0.01$M_{\odot}$ s$^{-1}$,
respectively.
} %%
\label{fig:a}
\end{figure}

Now we investigate the dependence of the energy spectrum of
anti-electron neutrino on the mass-accretion rate ($\dot{M}$). The
mass-accretion rate is theoretically not known while the numerical
simulations have been performed, because there are uncertainties on the
viscosity of the accretion disk and angular momentum of the
progenitor.  In Fig.~\ref{fig:a}, we show the energy spectrum for the
case of $\dot{M}$ = 0.01$M_{\odot}$ s$^{-1}$, 0.1$M_{\odot}$ s$^{-1}$,
1$M_{\odot}$ s$^{-1}$, 5$M_{\odot}$ s$^{-1}$, and 10$M_{\odot}$
s$^{-1}$.  The energy spectrum is represented as $E_{\anti{\nu}_e}^2
dn/dE_{\anti{\nu}_e}$ [MeV] because the cross section of the reaction
$p + \anti{\nu}_e \to n + e^+ $ is proportional to the square of the
energy. It is clearly shown that the energy spectrum becomes hard, and
intensity becomes large as the mass-accretion rate increases. This is
because the temperature at the inner region of the accretion disk
becomes higher.  Here we have to give some comments on the application
limit for the present accretion disk model in this study.  High energy
electron-positron pairs whose energies are greater than $\sim 100$MeV
produce muons and charged pions, which, in turn, decay into low energy
leptons. Such microphysics is not included in the present model.  The
influence might become important for the high accretion rate model.
In addition, as the density becomes large, the assumption that the
neutrinos are optically thin might become to break
down~\cite{popham99}. For example, the opacity of electron neutrino
($\tau$) at the inner most region of the accretion disk is estimated
as~\cite{bethe90}
\begin{eqnarray}
\tau &=&   \int \frac{1}{2}N_A \rho \sigma dr \\
%     &=& \int 3 \times 10^{33} \rho_{10} \times 9 \times 10^{-44} 
%         \epsilon^2 dr\\
%     &\sim& 1.6 \times 10^{-9} T_{\rm MeV}^2 \int \rho_{10} dr \\
     &\sim& 1.6 \times 10^{-9} T_{\rm MeV}^2  \rho_{10} H \\
     &\sim& 4.1 \times 10^{-3} \left( \frac{3 M_{\odot}}{M} \right) ^{1.5}
  \left( \frac{\dot{M}}{0.1 M_{\odot} s^{-1}} \right) ^{1.23}  
    \frac{1}{ \displaystyle{ \left(\frac{r}{r_s}\right)^{0.26}
%%    \\ \nonumber  &\times& \frac{1}{ \left(\frac{r}{r_s}\right)^{0.26}
          \left( 1 + \left( \frac{r}{r_s} \right)  \right)^{1.57}}
        },
\label{eq:opacity}
\end{eqnarray}
where $N_A$ is the Avogadro's number, $\sigma$ is the cross section of
the neutrino off a nucleon, $\rho_{10}$ is the density normalized by
10$^{10}$ g cm$^{-3}$. Therefore, $\tau$ may comparably become unity
if $\dot{M}$ is approximately $10 M_{\odot}$ s$^{-1}$, $r \sim r_g$,
and $M \sim 3 M_{\odot}$.  From the above considerations, models with
high accretion rate ($ \dot{M} \sim 10 M_{\odot}$ s$^{-1}$) may be the
extreme cases.
%In fact, Popham et al. (1999)
%concluded that the actual luminosity could be as much as a factor of
%5 lower when the effect of opacity is taken into account in their
%model of $\dot{M} = 10 M_{\odot}$ s$^{-1}$. 

\begin{figure}[htbp]
%\vspace{-0.8cm}
\centerline{\psfig{figure=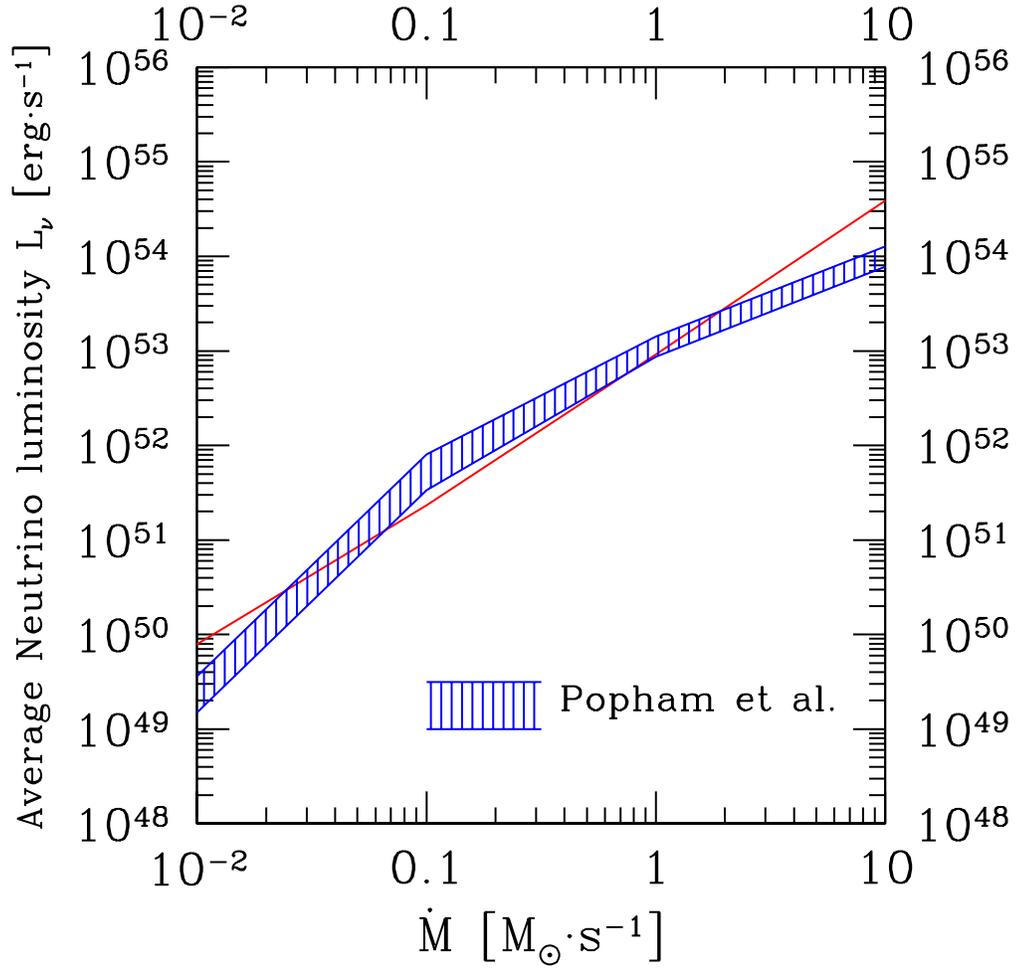,width=15cm}}
%\vspace{-0.5cm}
\caption{
Average neutrino luminosity as a function of $\dot{M}$.  We also show
the result of Popham et al. (1999) for comparison. }
\label{fig4}
\end{figure}

In Fig.~\ref{fig4}, we show the average neutrino luminosity which is
defined by
\begin{eqnarray}
    \label{eq:L_nu2}
    \anti{L}_{\nu} = Q_{\rm tot} / \Delta t_{\nu,{\rm emit}}.
\end{eqnarray}
For comparison, in the figure we also show the result of Popham et al.
(1999).  The uncertainty of their results comes from the Kerr
parameter of the central black hole ($a = 0$ -- 0.5).  From the
figure, we find that our simple formula reproduce their results fairly
well in wide range of mass-accretion rate. At high mass-accretion rate
($\dot{M} \sim 10 M_{\odot}$ s$^{-1}$), our formula may overestimate
the neutrino luminosity by a factor of three.  As shown below (see
section~\ref{results}), however, we also compute the GRB neutrino
background in such models as extreme cases.  In future, improvements
of the accretion model will enhance the predictability of the flux of
GRB neutrino background.

The event rates of $\anti{\nu}_e$ expected in water Cherenkov
detectors is represented by
\begin{equation}
    \label{eq:event}
    \frac{d R}{dE_{e^+}} =\frac{N_p}{4\pi D^2}
    \sigma_{p\anti{\nu}_e}(E_{e^+}) \frac{d n}{dE_{\anti{\nu}_e}}(E_{e^+}),
\end{equation}
where $E_{e^+} = E_{\anti{\nu}_e} - Q$ is the energy of the positron
which is scattered through $p + \anti{\nu}_e \rightarrow n + e^+$ in
the detector, $\sigma_{p\anti{\nu}_e}=
\frac{G_F^2}{\pi}(1+3g_A^2)E_{e^+}\sqrt{E_{e^+}^2-m_e^2}$ is the cross
section of the process, and $D$ is the distance from the Earth to the
collapsar. $N_p$ is the number of proton in the detector, e.g., $N_p
\simeq 1.5 \times 10^{33}$ for SK.
Thus the total event rate is
estimated by the integration,
\begin{eqnarray}
    \label{eq:R_eve}
    R = \int dE_{e^+}\frac{d R}{dE_{e^+}}.
\end{eqnarray}
%%
%We show just below that the
%total event rate is very sensitive to the mass-accretion rate (see
%Fig.~\ref{fig4}).
%In a similar fashion, we plot the total event rate as a function of
%$\dot{M}$ in Fig.~\ref{fig4}. In addition, we also estimate
%the total energy of $\anti{\nu}_e$ as a function of $\dot{M}$ by
%%%
%\begin{eqnarray}
%    \label{eq:tot_enrgy_mdot}
%    E_{\anti{\nu}_e,{\rm tot}} = \int \frac{d
%    n}{dE_{\anti{\nu}_e}}(E_{\anti{\nu}_e}) dE_{\anti{\nu}_e},
%\end{eqnarray}
%%%
%in Fig.~\ref{fig4}.  From the figure, approximately we can
%fit the total event rate from a single source as
%%
%\begin{eqnarray}
%    \label{eq:tot_event_mdot}
%    R \simeq 1.0 \left(\frac{\dot{M}}{0.1~
%      M_{\odot}~{\rm s}^{-1}}\right)^{1.7}\left(\frac{D}{3~{\rm
%      Mpc}}\right)^{-2}.Fig
%\end{eqnarray}

%%%%%%%%%%%%%%%%%%%%%%%%%%%%%%%%%%%%%%%%%%%%%%%%%%%
\subsubsection{Estimate for the total explosion energy}
\begin{figure}[htbp]
%\vspace{-0.8cm}
\centerline{\psfig{figure=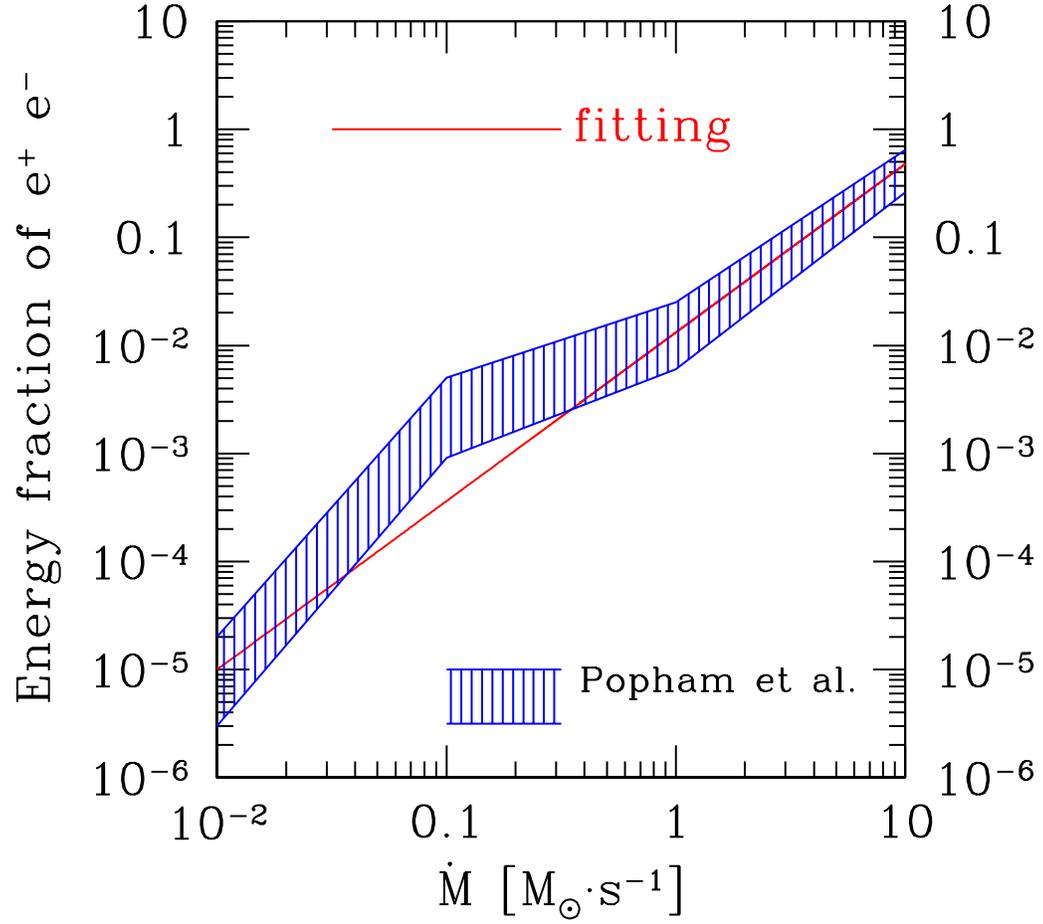,width=15cm}}
%\vspace{-0.5cm}
\caption{
Efficiency of $\nu_e \anti{\nu}_e$ pair annihilations as a function of
the mass-accretion rate.  For comparison, we also show the result of
Popham et al. (1999) as a shaded region.}
\label{fig6}
\end{figure}

\begin{figure}[htbp]
%\vspace{-0.8cm}
\centerline{\psfig{figure=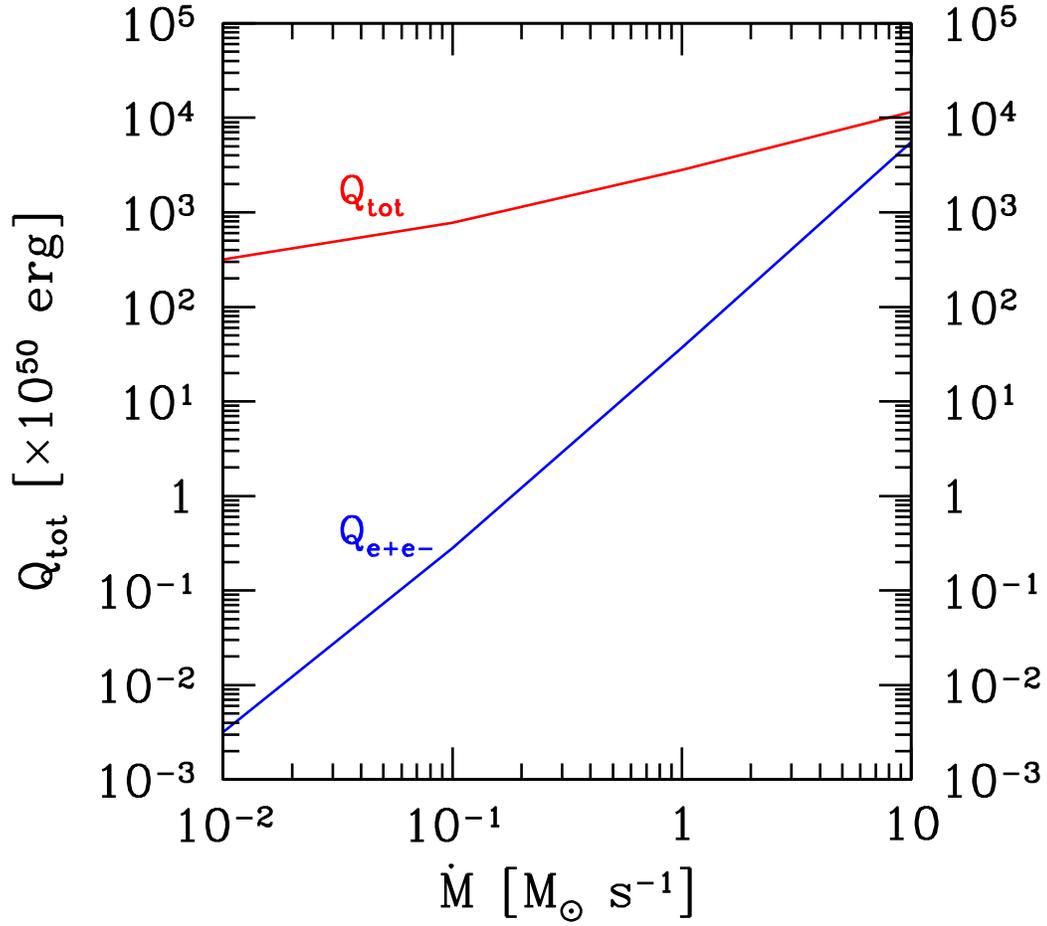,width=15cm}}
%\vspace{-0.5cm}
\caption{ %%
Total energy of emitted neutrinos ($Q_{\rm tot}$) and total deposited
energy ($Q_{e+e-}$) in units of erg as a function of the
mass-accretion rate.
}%%
\label{qtotqee}
\end{figure}

\begin{figure}[htbp]
%\vspace{-0.8cm}
\centerline{\psfig{figure=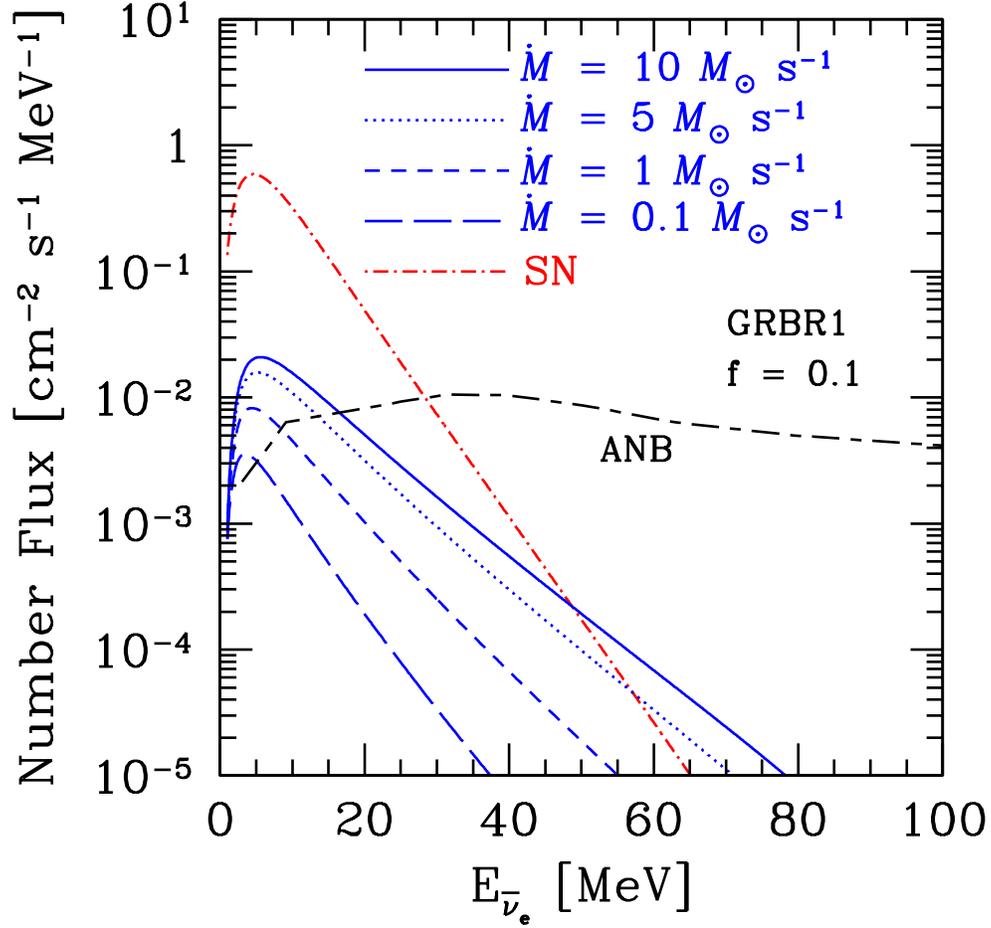,width=15cm}}
%\vspace{-0.5cm}
\caption{
Differential number flux of GRB neutrino background [cm$^{-2}$
s$^{-1}$ MeV$^{-1}$].  Solid line, dotted line, short-dashed line, and
long-dashed line correspond to $\dot{M}$ = 10$M_{\odot}$ s$^{-1}$,
5$M_{\odot}$ s$^{-1}$, 1$M_{\odot}$ s$^{-1}$, and 0.1$M_{\odot}$
s$^{-1}$, respectively. The case of SN neutrino background 
and atmospheric neutrino background (ANB) are shown in dot-short-dashed
line and long-short-dashed line for comparison.
}
\label{flux1}
\end{figure}

\begin{figure}[htbp]
%\vspace{-0.8cm}
\centerline{\psfig{figure=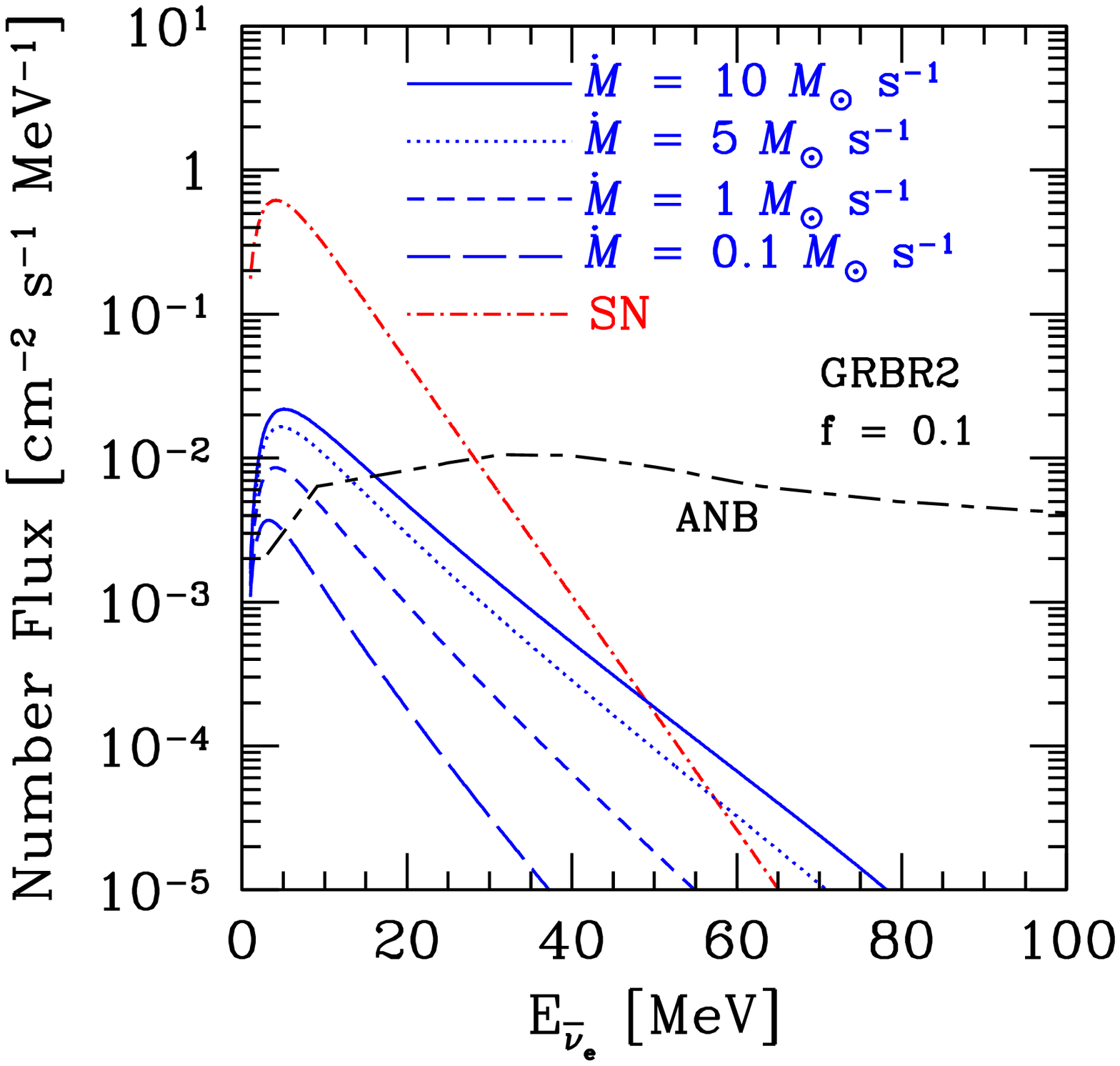,width=15cm}}
%\vspace{-0.5cm}
\caption{
Same as Fig.9 but GRBR2 is used.
}
\label{flux2}
\end{figure}

\begin{figure}[htbp]
%\vspace{-0.8cm}
\centerline{\psfig{figure=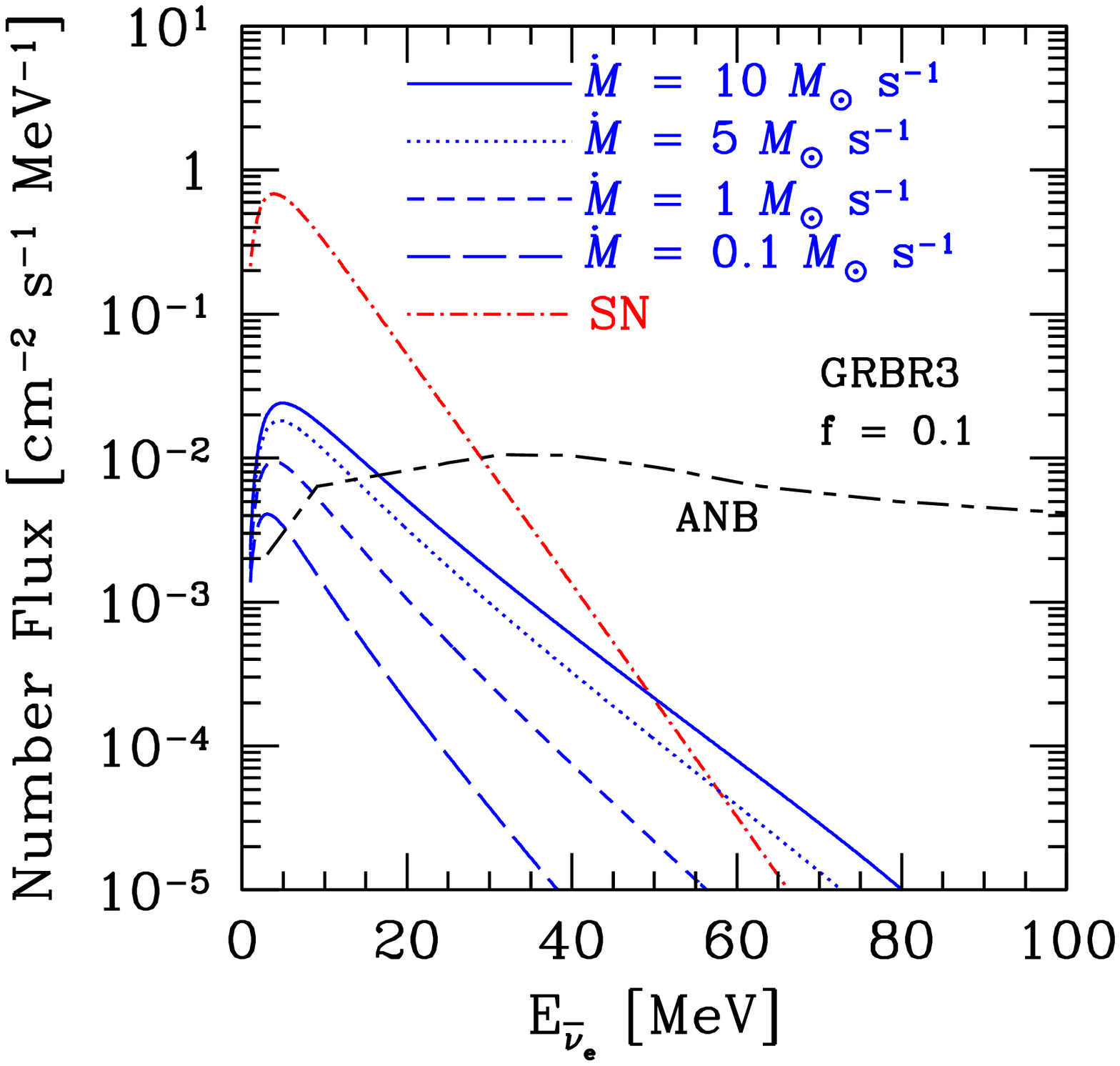,width=15cm}}
%\vspace{-0.5cm}
\caption{
Same as Fig.9 but GRBR3 is used.
}
\label{flux3}
\end{figure}

%%%%%%%%%%%%%%%%%%%%%%%%%%%%%%%%%%%%%%%%%%%%%%%%%%%
It is reported that the energy deposition rate becomes as large as
$\sim 10^{51}$ erg s$^{-1}$ by the $\nu \anti{\nu}$ pair
annihilation~\cite{popham99}. In their studies, they computed the
energy deposition rate considering the shape of the accretion
disk. Their results are shown in Fig.~\ref{fig6} as a shaded
region. Here the efficiency of $\nu_e \anti{\nu}_e$ pair annihilations
is defined as $\epsilon \equiv Q_{e+e-}/Q_{\rm tot}$, where $Q_{e+e-}$
is the total deposited energy.  Their results are fitted well as
\begin{eqnarray}
\epsilon = 1.32 \times 10^{-2} \left(
\frac{\dot{M}}{1 M_{\odot} s^{-1}} \right)^{1.56},
\label{efficiency}
\end{eqnarray} 
which is also shown in Fig.~\ref{fig6}. In this study, we use this
fitting formula to estimate the explosion energy of a GRB. 
%As for $Q_{\rm tot}$, it can be obtained as
%\begin{eqnarray}
%$
%Q_{\rm tot} = 2 \int  E_{\anti{\nu}_e}  dn/dE_{\anti{\nu}_e} d
%E_{\anti{\nu}_e},
%$
%\label{eq:qtot}
%\end{eqnarray} 
%where the total energy of electron neutrinos is assumed to be equal
%to that of anti-electron neutrinos. 
Then, we obtain the total deposited energy by $Q_{e^+e^-}$ = $\epsilon
Q_{\rm tot}$.  In Fig.~\ref{qtotqee}, the total energy of emitted
neutrinos ($Q_{\rm tot}$) and the total deposited energy ($Q_{e+e-}$)
are shown as a function of the mass-accretion rate. Since the total
explosion energy of SN1998bw is reported to be $\sim 10^{52}$
erg~\cite{macfadyen99}, we find that it corresponds to the
mass-accretion rate $\sim (1-2) M_{\odot} s^{-1}$ in
Fig.~\ref{qtotqee}.

For simplicity, we ignored the heating process from $\nu_{\mu}
\anti{\nu}_{\mu}$ and $\nu_{\tau} \anti{\nu}_{\tau}$ pair
annihilations in this computation because the effect is much smaller
than $\nu_{e} \anti{\nu}_e$ pair annihilation. The reason is as
follows. In the electromagnetic thermal bath, $\nu_{\mu}
\anti{\nu}_{\mu}$ and $\nu_{\tau} \anti{\nu}_{\tau}$ pairs are
produced from $e^+e^-$ pair annihilations only through the neutral
current interaction. In addition, $\nu_{\mu} \anti{\nu}_{\mu}$ and
$\nu_{\tau} \anti{\nu}_{\tau}$ pairs annihilate into $e^+e^-$ pairs
only through the neutral current interaction. On the other hand
$\nu_{e} \anti{\nu}_e$ pairs are produced through both the charged
and neutral currents and annihilate into $e^+e^-$ pairs
correspondingly through the same process. Then, compared with the case
of $\nu_{e} \anti{\nu}_e$, the efficiency of the heating from
$\nu_{\mu} \anti{\nu}_{\mu}$ and $\nu_{\tau} \anti{\nu}_{\tau}$ pairs
in the electromagnetic thermal plasma is just within $10\%$ of the
total, even if the $\nu\anti{\nu}$ pair production from $e^+e^-$ pair
annihilation dominates all the other processes of neutrino production.

\subsection{GRB neutrino background}
%%%%%%%%%%%%%%%%%%%%%%%%%%%%%%%%%%%%%%%%%%%%%%%%%%%
In order to get the differential number flux of back ground neutrinos,
first we compute the present number density of the background
neutrinos per unit neutrino energy, $dn (E_{\anti{\nu}_e}) /
dE_{\anti{\nu}_e}$ in Eq.~(\ref{eq:dndE}). The contribution of the
neutrinos emitted in the interval of the redshift $z \sim z +dz$ is
given as
\begin{eqnarray}
    dN_{\nu} (E_{\anti{\nu}_e}) = R_{\rm GRB} (z) \frac{dt}{dz} dz
    \frac{dn (E_{\anti{\nu}_e}')}{dE_{\anti{\nu}_e}'} (1+z)
    dE_{\anti{\nu}_e},
    \label{eq:a}
\end{eqnarray} 
where $E_{\anti{\nu}_e}' = (1+z)E_{\anti{\nu}_e}$.  It is noted that
$R_{\rm GRB}$ is GRB rate per comoving volume, and the effect of the
expansion of the universe is taken into account in
Eq.~(\ref{eq:a}). The Friedmann equation gives the relation between
$t$ and $z$ as follows:
\begin{eqnarray}
        \frac{dz}{dt} =
%        \frac{dz}{dt} &=& \\ \nonumber
  &-& H_0 (1+z) \sqrt{
  (1+\Omega_M z) (1+z)^2 - \Omega_{\Lambda} (2z +z^2) }.
    \label{eq:b}
\end{eqnarray} 
We now obtain the differential number flux of GRB, $dF_{\nu}
(E_{\anti{\nu}_e}) / dE_{\anti{\nu}_e}$, by using the relation $dF_{\nu}
(E_{\anti{\nu}_e})/dE_{\anti{\nu}_e} = c dN_{\nu} (E_{\anti{\nu}_e}) /
dE_{\anti{\nu}_e}$:
\begin{eqnarray}
    \frac{dF_{\nu}}{dE_{\anti{\nu}_e}} = &&\frac{c}{H_0}
    \int^{z_{\rm max}}_{0} R_{\rm GRB} (z) \frac{dn
    ((1+z)E_{\anti{\nu}_e}) } {dE_{\anti{\nu}_e}}
%        \\ \nonumber  &&\times 
        \frac{dz}{ \sqrt{(1+ \Omega_M z) (1+z)^2 - \Omega_{\Lambda}
        (2z +z^2)  }}
    \label{eq:c}
\end{eqnarray}

%%%%%%%%%%%%%%%%%%%%%%%%%%%%%%%%%%%%%%%%%
\section{Results}\label{results}
%%%%%%%%%%%%%%%%%%%%%%%%%%%%%%%%%%%%%%%%%
%back ground$B$r$$$m$s$J%Q%i%a!<%?$K$D$$$F!#(B

The differential number flux of GRB neutrinos is shown in
Figs.~\ref{flux1},~\ref{flux2} and~\ref{flux3}. As for the GRB
formation history, GRBR1, GRBR2, and GRBR3 are assumed,
respectively. It is found that the differential number flux is similar
with each other even if the different star formation history is
adopted. That is because the energy of neutrinos from high-z GRBs is
redshifted as $\propto 1/(1+z)$, and the volume containing high-z GRBs
is relatively small $ \propto 1/\sqrt{(1+ \Omega_M z) (1+z)^2 -
\Omega_{\Lambda} (2z +z^2)}$ as given in Eq.~(\ref{eq:c}). Thus, we
show the results only for GRBR3 below.  On the other hand, we can see
clearly that the differential number flux becomes larger as the
mass-accretion rate increases. In particular, in high accretion model
such as $\dot{M} = 5M_{\odot}$, the intensity of the GRB neutrino
background is larger than that of SN neutrino background in the energy
region $E_{\anti{\nu}_e} \ge 50$MeV.  As discussed below, in such a
high energy region, the neutrino background will be obscured by the
noise from the atmospheric neutrino background (ANB). As the event
number increases, however, we will have statistically sufficient data,
and the signals from GRBs will dominate the statistical error of
ANB. In the next section we estimate the time required to detect the
signals from GRB (see also Table~\ref{tab:table2}).  Note that at the
energy which is larger than $50$MeV, the signals from GRBs dominate
the noise of electrons from the decay of invisible muons because their
spectrum has a steep cut-off at $50$MeV.  In the next section, we will
discuss them in detail.

\begin{figure}[htbp]
%\vspace{-0.8cm}
\centerline{\psfig{figure=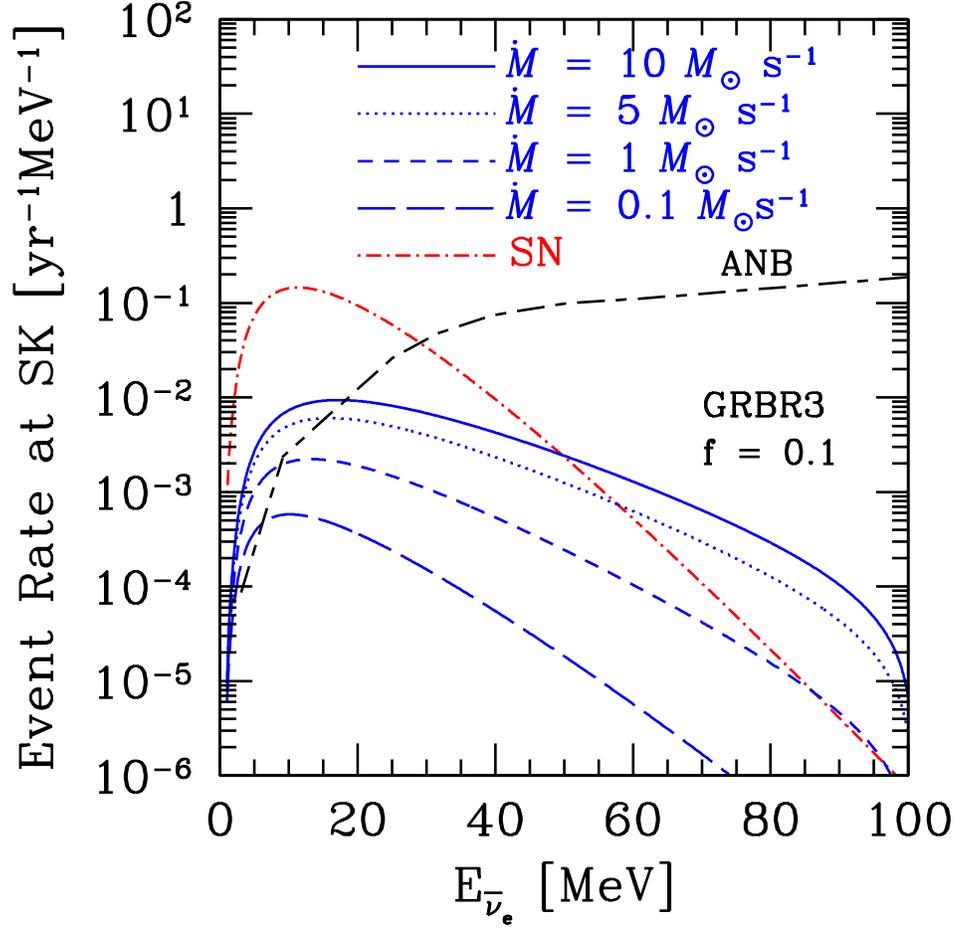,width=15cm}}
%\vspace{-0.5cm}
\caption{
Event rate of GRB neutrino background
[yr$^{-1}$ MeV$^{-1}$] at Super-Kamiokande.
Solid line, dotted line, short-dashed line, and long-dashed line
correspond to $\dot{M}$ = 10$M_{\odot}$ s$^{-1}$, 5$M_{\odot}$ s$^{-1}$,
1$M_{\odot}$ s$^{-1}$, and 0.1$M_{\odot}$ s$^{-1}$,
respectively. Event rate of SN neutrino background and atmospheric neutrino
background (ANB) are shown in dot-short-dashed line and long-short-dashed
line for comparison.
}
\label{event1}
\end{figure}

The event rates at SK are shown in Fig.~\ref{event1} in units of
yr$^{-1}$ MeV$^{-1}$ for various mass-accretion rates. For event
rates, increasingly we cannot discriminate among the three models of
the star formation history because the cross section of $p +
\anti{\nu}_e \rightarrow n + e^+$ is proportional to the square of the
neutrino energy ($\propto E_{\anti{\nu}_e}^2$), and the relative
contribution of low energy neutrinos to the event rate is
small. Therefore, irrespective of the differences among the models of
the star formation history, we can expect the event rate of GRB
neutrino background as a function of mass-accretion rate. Namely the
informations on the average mass-accretion rate will be obtained from
such an observation, which, in turn, should give us informations on
the total explosion energy of a GRB (see Fig.~\ref{qtotqee}). They
are, of course, never obtained only by the gamma-ray observations.

\begin{figure}[htbp]
%\vspace{-0.8cm}
\centerline{\psfig{figure=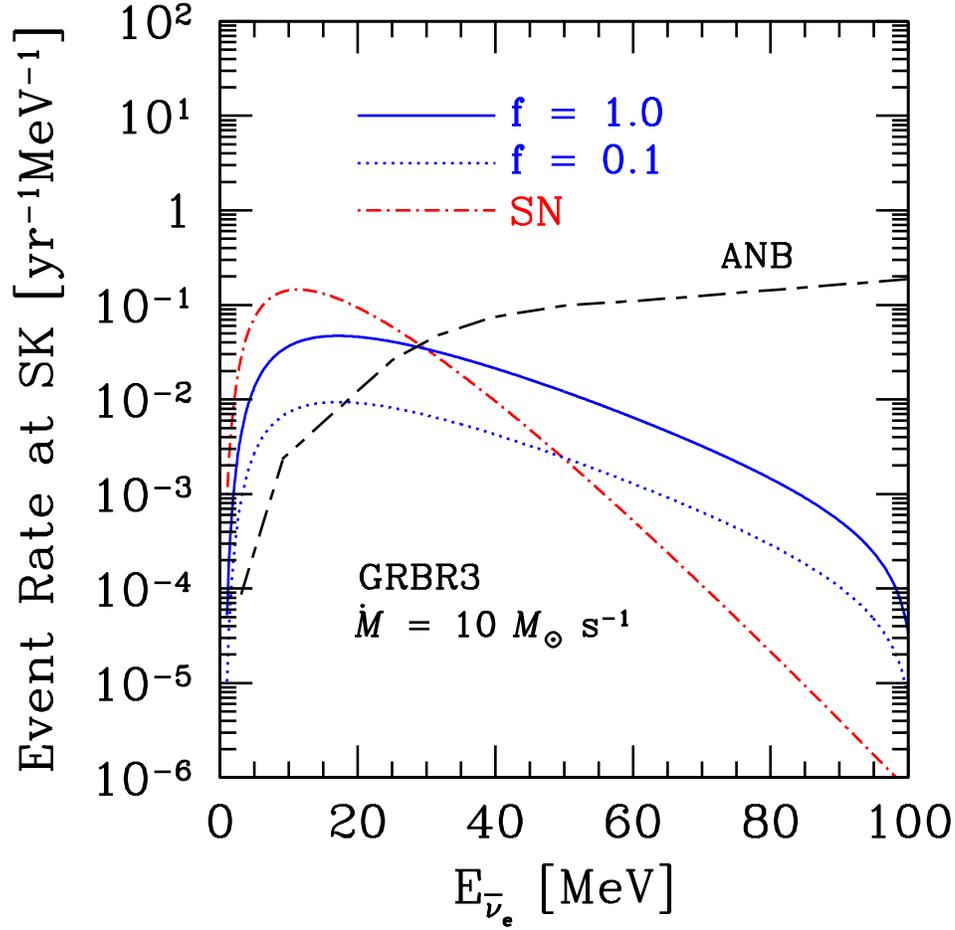,width=15cm}}
%\vspace{-0.5cm}
\caption{ %%
Dependence of the event rates on the probability $f$ that one
collapsar generates a GRB.  Solid line and dotted line correspond to
the case $f=1.0$ and $f=0.1$, respectively.  We adopt the case of
$\dot{M} = 10 M_{\odot} \s^{-1}$
}%%
\label{event2}
\end{figure}

In Fig.~\ref{event2}, we show the dependence of the event rates on the
probability $f$ that one collapsar generates a GRB in the case of
$\dot{M} = 10 M_{\odot}$ s$^{-1}$. It is clear that the signals from
GRBs dominate those from SN for $E_{\bar{\nu}_e} \ge 30$MeV in the
case of $f=1.0$. However, the noise from the atmospheric neutrinos
obscures the signals of GRBs even if we adopt the high value of $f$
(=1.0).  In the next section, we discuss the detectability of the
signals of the GRB neutrino background as a significant deviation from
the statistical error of ANB.

To estimate the event rates more realistically, we should consider the
effects of neutrino oscillation. The flavor eigen states
$\nu_{\alpha}$ ($\alpha, = e, \mu,$ and $\tau$) are related with the
mass eigen states $\nu_i$ (``$i$'' = 1, 2, and 3) by the mixing matrix
$U$ as $\nu_{\alpha} = \sum_i U_{\alpha i}\nu_i$. Then, we
parameterize it by
\begin{eqnarray}
    \label{eq:mixing_gen}
    U = \left( %%
      \begin{array}{ccc}
          c_{13}c_{12} &
          c_{13}s_{12} & s_{13} \\ %%
          -c_{23} s_{12} -s_{23}s_{13} c_{12} & c_{23} c_{12} - s_{23}
          s_{13} s_{12} &
          s_{23} c_{13} \\ %%
          s_{23}s_{12} - c_{23} s_{13} c_{12} & -s_{23}c_{12} -
          c_{23}s_{13}s_{12} & c_{23}c_{13}
      \end{array} 
    \right), %%
\end{eqnarray}
with $s_{ij} = \sin\theta_{ij}$ and $c_{ij} = \cos\theta_{ij}$, where
we ignored the CP phase in the lepton sector. In general, the
probability of the vacuum oscillation from a flavor $\nu_{\alpha}$ to
a flavor $\nu_{\beta}$ is represented by
\begin{eqnarray}
    \label{eq:prob1}
    P(\nu_{\alpha} \to \nu_{\beta}) =
    \langle\nu_{\beta}|\nu_{\alpha}(t)\rangle = \sum_i |U_{\beta i}|^2
    |U_{\alpha i}|^2 + 2 Re \sum_{i > j} U_{\beta i} U_{\beta j}^*
    U_{\alpha i}^* U_{\alpha j} e^{-i \frac{L}{2E_{\nu}} \delta m^2_{i
    j}},
\end{eqnarray}
where $L$ is the distance from the source to the Earth (= $ct$),
$E_{\nu}$ is the neutrino energy, and the mass difference is defined
by $\delta m^2_{i j} \equiv m_j^2 - m_i^2$. For simplicity, here we
assume the normal hierarchy of neutrino masses, $m_1 < m_2 < m_3$. The
results of recent oscillation experiments are as follows. The solar
neutrino data favor the Large Mixing Angle solution ($0.6 \le
\sin^22\theta_{12} \le 0.98$), and $2\times 10^{-5} \ev^2 \le \delta
m^2_{12} \le 4\times 10^{-4} \ev^2$ at 3$\sigma$
C.L.~\cite{Bahcall:2001cb}.  The atmospheric neutrino data also imply
the large mixing ($\sin^22 \theta_{23} \ge 0.85$), and $1.1\times
10^{-3} \ev^2 \le \delta m^2_{23} ( \simeq \delta m^2_{13}) \le 5
\times 10^{-3} \ev^2$ at 99 $\%$ C.L.~\cite{Toshito:2001dk}. The CHOOZ
reactor experiment shows that $\sin^22\theta_{13} \le 0.1$ at 95 $\%$
C.L~\cite{Apollonio:1999ae}. Then, the above data tells us that
$\delta m^2_{12} \ll \delta m^2_{23} \sim \delta m^2_{13}$.

Adopting the above data, we can assume that the mixing matrix is
approximately simplified into
\begin{eqnarray}
    \label{eq:mixing}
    U \simeq \left( %%
      \begin{array}{ccc}
          \sqrt{2}/2 & \sqrt{2}/2 & 0 \\%%
          - 1/2 & 1/2 & \sqrt{2}/2 \\%%
          1/2 & - 1/2 & \sqrt{2}/2
      \end{array}
    \right). %
\end{eqnarray}
Because neutrinos propagate for long distances, the second term in
Eq.~(\ref{eq:prob1}) is averaged and becomes zero. Then, we have
\begin{eqnarray}
    \label{eq:nuebar1}
    P({\anti{\nu}_e \to \anti{\nu}_e}) &=& 1/2, \\%%
    \label{eq:nuebar2}
    P({\anti{\nu}_{\mu} \to \anti{\nu}_e}) &=& 1/4, \\%%
    \label{eq:nuebar3}
    P({\anti{\nu}_{\tau} \to \anti{\nu}_e)} &=& 1/4.
\end{eqnarray}
If we choose the inverted hierarchy, i.e., $\delta m^2_{32} \ll \delta
m^2_{21} \simeq \delta m^2_{31}$, we only interchange the roles of
$m_1$ and $m_3$ in Eq.~(\ref{eq:mixing_gen}). Namely we only
interchange the first and third columns of $U$. Even then, it is
remarkable that the results in
Eq.~(\ref{eq:nuebar1}),~(\ref{eq:nuebar2}), and~(\ref{eq:nuebar3}) are
not changed at all.

\begin{figure}[htbp]
%\vspace{-0.8cm}
\centerline{\psfig{figure=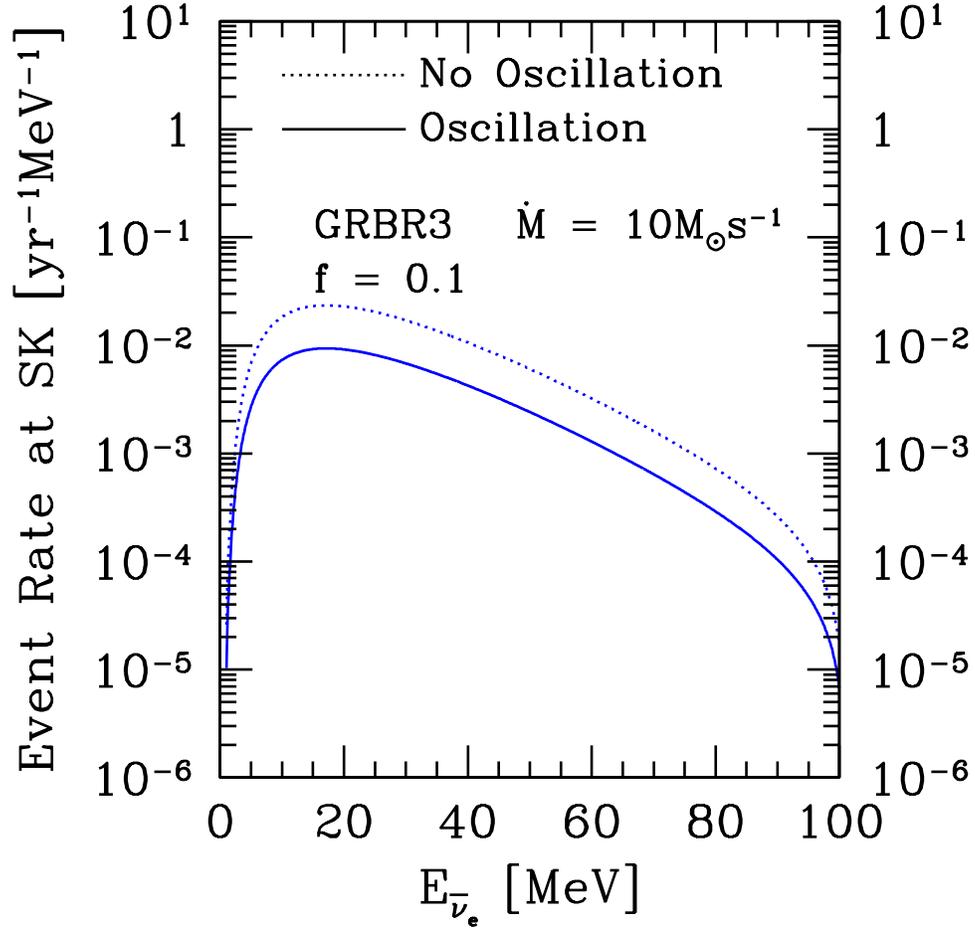,width=15cm}}
%\vspace{-0.5cm}
\caption{ %%
Event rate at SK for the model of $\dot{M} = 10M_{\odot}$.  Solid
lines represent the case in which the effects of neutrino oscillation
are taken into consideration. Dotted lines represent the case of no
oscillation.
} %%
\label{osc}
\end{figure}

In Fig.~\ref{osc}, we show the event rate as a function of
$E_{\anti{\nu}_e}$ at SK in which we consider the effects of the
neutrino oscillation for $\dot{M} = 10M_{\odot}$.  It is noted that
the flux of $\anti{\nu}_e$ becomes lower as a result of the neutrino
oscillation. That is because $\anti{\nu}_{\mu}$ and
$\anti{\nu}_{\tau}$ are produced only through the $e^+ e^-$ pair
annihilation.

%%%%%%%%%%%%%%%%%%%%%%%%%%%%%%%%%%%%%%%%%%%%%%%%%%%
\section{Discussions}\label{discussion}
%%%%%%%%%%%%%%%%%%%%%%%%%%%%%%%%%%%%%%%%%%%%%%%%%%%
% $BF@$i$l$?>pJs$N0U5A$K$D$$$F!#$^$@$^$@$D$1$?$7!#(B
%At present, the
%accretion disk model is too simple and there are much uncertainties
%on this model (see Fig.~\ref{fig4}). Thus we can not determine
%the total explosion energy of a GRB precisely at present even if the
%GRB neutrino background is detected, although our conclusion
%that GRB neutrino background will be detected as long as the
%average mass-accretion rate is high ($\sim$ a few $M_{\odot}$ s$^{-1}$)
%wil be correct quantitatively.
%By including and investigating the microphysics in the accretion disk,
%more precise model should be obtained and total explosion energy will
%be obtained more correctly.

% $BB>$N(Bbackground$B$K$D$$$F$N5DO@(B($BBg5$!"(Binvisible etc)

We consider the detectability of the neutrino background from GRBs.
In the energy range $E \le 50$MeV, the neutrino background is obscured
by the events of the electrons which come from the decay of invisible
muons~\cite{nakahata02}.
The spectrum of the electrons is given by $ dN/dE_e = (G_F^2/12
\pi^3) m_{\mu}^2 E_e^2 (3 - 4E_e/m_{\mu}), $ where $m_{\mu}$ is the
muon mass, and $E_e$ is the electron (positron) energy.  The event
rate of the electrons from the invisible muons at SK is about $\sim$ 1
in units of yr$^{-1}$ MeV$^{-1}$ in the energy range $50 \ge
E_{\anti{\nu}_e} \ge 20$MeV~\cite{ando02}. Thus, signals of GRB
neutrino background are smaller than those of electrons from invisible
muons in the range. In addition, the signals from SN neutrino
background will basically dominate those of GRB neutrino background in
the energy range. On the other hand, the spectrum of electrons from
the invisible muons has a steep cut-off at about 50 MeV because the
maximum energy of the emitted electrons is about half of the muon
mass.  Thus, as mentioned above, the signal from GRBs becomes larger
than the noise from invisible muons in the energy range
$E_{\anti{\nu}_e} \ge 50$MeV.  In the case of the high mass-accretion
rate such as $\dot{M} \gtrsim 5M_{\odot}$, the signals of the GRB
neutrino background are larger than those of SN neutrino background in
the energy region.  However, the severest noise comes from the
atmospheric neutrinos. In Table~\ref{tab:table0}, we show the event
rate (yr$^{-1}$) of ANB and its statistical error in SK/TITAND in the
energy regions $E_{\bar{\nu}_e}$ = (50~--~60)MeV and $E_{\bar{\nu}_e}
\ge$ 50MeV~\cite{gaisser88,honda95}.  TITAND is proposed as a
next-generation plan of the multi-megaton water Cherenkov
detector~\cite{suzuki:2001}. The first phase of TITAND is planed to be
with 2 Mt water inside. Thus, the event rate becomes approximately one
hundred times larger than SK.

For comparison, we show the event rates from the GRB neutrino
background at SK/TITAND in Table~\ref{tab:table1} for various
models. The third column of the table represents whether we consider
the effects of the neutrino oscillation or not. 'Yes' means that the
effects of the neutrino oscillation are taken into account. It is
apparent that the signals from GRB are smaller than the noise from
ANB. As the event number increases, however, we will have
statistically sufficient data, and the signals from GRBs will dominate
the statistical error of ANB. Namely, then we can definitely detect
the signals of the GRB neutrino background as a significant deviation
from the statistical error of ANB. The time (yr) required to
marginally distinguish the signals from the noise is computed by $t =
N/S^2$, where $N$ is the noise (yr$^{-1}$) from ANB and $S$ is the
signals from GRBs (yr$^{-1}$). In Table~\ref{tab:table2}, we show the
time required to detect the signal of GRBs at SK/TITAND for the energy
range $E$ = (50~--~60) MeV and $E \ge 50$ MeV.  From the table, we
find that we can detect the signals within $\sim$10 years at TITAND in
the case that the mass-accretion rate is (5~--~10$M_{\odot}$ s$^{-1}$)
and $f$ is (0.5~--~1).
Although it seems to take a long time to detect the
signals from GRBs for the cases in which the conservative parameter
sets are adopted, constraints for optimistic parameter sets will be
obtained by the observations of TITAND.
Of course, a larger water Cherenkov
detector than TITAND will detect the signals from GRBs more
immediately even if the GRBs obey the conservative parameters.

%From Figs.~\ref{event1}-~\ref{event3},
%the mass-accretion rate is concluded to be lower than
%$\sim 4 M_{\odot}$ s$^{-1}$ for our model.
%This corresponds to the total
%explosion energy of a GRB is lower than $\sim 10^{52}$ erg.
%This will be the most severe constraint for the total explosion
%energy of a GRB.

% GBR = Collapsar?
In this study, we have mainly assumed that one tenth of collapsars
whose mass range are in (35~--~125)$M_{\odot}$ are accompanied with
GRBs. Here we investigate the validity of the assumption. The locally
observed value for the SN rate is in the range $R_{\rm SN} = (1.4 \pm
0.5) \times 10^{-4} h_{0.7}^3$ yr$^{-1}$ Mpc$^{-3}$, where $h =
H_0/$(100 km s$^{-1}$ Mpc$^{-1}$)~\cite{madau98}.  On the other hand,
it is reported that the GRB rate at $z=0$ is $2.4 \times 10^{-9}
h_{0.7}^3$ yr$^{-1}$ Mpc$^{-3}$ using the BATSE data
catalog~\cite{totani97}. On the other hand, we assumed that the ratio
of the collapsar rate to SN rate is $\sim 0.13$ using
Eqs.~(\ref{eq:4}) and~(\ref{eq:4.2}).  Thus, we estimate that the
collapsar rate at $z=0$ is $1.8 \times 10^{-5}  h_{0.7}^3$ yr$^{-1}$
Mpc$^{-3}$. Adopting the beaming factor is $d \Omega \sim 4 \pi \times
10^{-3}$, and the probability that one collapsar generates a GRB is $f
= 0.1$, we estimate the GRB rate at $z=0$:
\begin{eqnarray}
R_{\rm GRB} = 1.8 \times 10^{-9} \left( \frac{f}{0.1}   \right)
              \left( \frac{d \Omega}{4 \pi \times 10^{-3}}   \right)
              \left( \frac{h}{0.7}   \right)^3 ,
    \label{eq:rate}
\end{eqnarray}
in units of $\rm yr^{-1} \ Mpc^{-3}$ which is consistent with the
previous work. At present, the average beaming angle is not determined
precisely except for some observations that suggest it is several
degree~\cite{frail01}.  Therefore, we leave the value of $f$ as a free
parameter at present.  In addition, the previous works on the GRB
formation rate would have a lot of uncertainties since only the BATSE
data catalog is used in these analyses, and the average luminosity of
GRBs is adopted to fit the model.  More precise data on the beaming
angle and on the GRB formation rate (for example, by using the
peak-lag relation~\cite{norris00}) will give us informations on the
probability $f$. Then, our predictions of
GRB neutrino background become more correct.

In this study, we have assumed that GRB formation rate is proportional
to the star formation rate in the universe. If this assumption is not
correct, of course, the resulting energy spectrum will also be
changed. Therefore, on the contrary, when we observe the differential
flux of GRB neutrino background, we will also be able to know whether
the GRB formation history traces the star formation history or not.

We used the model presented by Popham et al. (1999) as an accretion
model. We have to investigate the model dependence of the neutrino
luminosity in future. In particular, we have to investigate the
neutrino-dominated accretion disk (NDAF) for the high accretion rate
cases~\cite{ref:kohri02, narayan01}.

% Neutrino$B?6F0$K$D$$$F!#(B(e+e-$B$,8z$-=P$9$+$b$M!#!#!#(B)
In this study, we have considered the effects of vacuum neutrino
oscillations. However, the effects of MSW effects are not taken
into account. As long as the collapsar model is adopted, the density
profile of the progenitor should depend on the zenith angle and time.
We are planning to investigate these effects in the near future.

% B-Z$B$G$b(Bneutrino$B$O$G$k$1$I!#!#!#$=$NJU$K$D$$$F!#(B
When the mechanism of the central engine is not the neutrino heating
but the Blandford-Znajek mechanism~\cite{blandford77}, the
observations of GRB neutrino background will give a lower limit of the
total explosion energy of a GRB.

%%%%%%%%%%%%%%%%%%%%%%%%%%%%%%%%%%%%%%%%%%%%%%%%%
\section{Summary and Conclusion}\label{summary}
%%%%%%%%%%%%%%%%%%%%%%%%%%%%%%%%%%%%%%%%%%%%%%%%%
We have estimated the flux of the GRB neutrino background and computed
the event rate at SK/TITAND in the collapsar models, assuming that GRB
formation history traces the star formation history. We have found
that the flux and the event rate depend sensitively on the
mass-accretion rate although the detection of signals from GRBs seems
to be difficult by SK.  On the other hand, however, we have found that
the GRB neutrino background may be detected by TITAND within $\sim$ 10
yrs as long as the average mass-accretion rate is high ($\gtrsim$ a
few $M_{\odot}$ s$^{-1}$), and the probability that one collapsar
generates a GRB is high ($f$ = 0.5~--~1.0). 
%A larger water Cherenkov
%detector than TITAND will detect the signals from GRBs more
%immediately even if the GRBs obey the conservative parameters. 
Thus, we conclude that the informations on the mass-accretion rate of
collapsar will be obtained by the observations of the GRB neutrino
background, which in turn, should give us informations on the total
explosion energy of GRBs in future.  They are, of course, never
obtained only by the gamma-ray observations. Although there are some
simplifications in our analyses and some uncertainties on the present
observations, our proposal to determine the total explosion energy of
a GRB is very challenging and interesting.  This paper is the maiden
attempt about the GRB neutrino background.  We will revise our model
to enhance the predictability, and our suggestions will be tested by
the future observations.

%%%%%%%%%%%%%%%%%%%%%%%%%%%%%%%%%%%%%%%%%%%%%%%%%%%%%%%%%%%%%%%%%%%%%
\acknowledgements
%%%%%%%%%%%%%%%%%%%%%%%%%%%%%%%%%%%%%%%%%%%%%%%%%%%%%%%%%%%%%%%%%%%%%
S.N. is grateful to R. Narayan for useful comments.
K.K. is grateful to T. Onogi for useful comments.
S.N. and K.K. thank the Yukawa Institute for Theoretical Physics at
Kyoto University, where this work was initiated during the YITP-W-01-06 on
'GRB2001' and completed during the YITP-W-99-99
on 'Blackholes, Gravitational Lens, and Gamma-Ray Bursts'.
The authors are grateful to Y. Totsuka, T.Suzuki, M. Nakahata, Y. Fukuda,
and A. Suzuki for useful comments on the neutrino background at
Super-Kamiokande.
This research has been supported in part by a Grant-in-Aid for the
Center-of-Excellence (COE) Research (07CE2002) and for the Scientific
Research Fund (199908802) of the Ministry of Education, Culture,
Sports, Science and Technology and by Japan Society for the Promotion of
Science Postdoctoral Fellowships for the Research Abroad.

\vspace{-0.5cm}

%%%%%%%%%%%%%%%%%%%%%%%%%%%%%%%%%%%%%%%%%%%%%%%%%%%%%%%%%%%%%%%%%%%%%%

%\end{document}

%%%%%%%%%%%%%%%%%%%%%%%%%%%%%%%%%%%%%%%%%%%%%%%%%%%%%%%%%%%%%%%%%%%%%%
%%%%%%%%%%%%%%%%%%%%%% FIGURES %%%%%%%%%%%%%%%%%%%%%%%%%%%%%%%%%%%%%%%
%%%%%%%%%%%%%%%%%%%%%%%%%%%%%%%%%%%%%%%%%%%%%%%%%%%%%%%%%%%%%%%%%%%%%%
\newpage

%%%%%%%%%%%%%%%%%%%%%%%%%%%%%%%%%%%%%%%%%%%%%%%%%%%%%%%%%%%%%%%%%%%%%%
\newpage

%%%%%%%%%%%%%%%%%%%%%%%%%%%%%%%%%%%%%%%%%%%%%%%%%%%%%%%%%%%%%%%%%%%%%%
\newpage

%%%%%%%%%%%%%%%%%%%%%%%%%%%%%%%%%%%%%%%%%%%%%%%%%%%%%%%%%%%%%%%%%%%%%%
\newpage

%%%%%%%%%%%%%%%%%%%%%%%%%%%%%%%%%%%%%%%%%%%%%%%%%%%%%%%%%%%%%%%%%%%%%%
\newpage

%%%%%%%%%%%%%%%%%%%%%%%%%%%%%%%%%%%%%%%%%%%%%%%%%%%%%%%%%%%%%%%%%%%%%%
\newpage

%%%%%%%%%%%%%%%%%%%%%%%%%%%%%%%%%%%%%%%%%%%%%%%%%%%%%%%%%%%%%%%%%%%%%%
\newpage

%%%%%%%%%%%%%%%%%%%%%%%%%%%%%%%%%%%%%%%%%%%%%%%%%%%%%%%%%%%%%%%%%%%%%%
\newpage

%%%%%%%%%%%%%%%%%%%%%%%%%%%%%%%%%%%%%%%%%%%%%%%%%%%%%%%%%%%%%%%%%%%%%%
\newpage

%%%%%%%%%%%%%%%%%%%%%%%%%%%%%%%%%%%%%%%%%%%%%%%%%%%%%%%%%%%%%%%%%%%%%%
\newpage

%%%%%%%%%%%%%%%%%%%%%%%%%%%%%%%%%%%%%%%%%%%%%%%%%%%%%%%%%%%%%%%%%%%%%%
\newpage

%%%%%%%%%%%%%%%%%%%%%%%%%%%%%%%%%%%%%%%%%%%%%%%%%%%%%%%%%%%%%%%%%%%%%%
\newpage

%%%%%%%%%%%%%%%%%%%%%%%%%%%%%%%%%%%%%%%%%%%%%%%%%%%%%%%%%%%%%%%%%%%%%%
\newpage

%%%%%%%%%%%%%%%%%%%%%%%%%%%%%%%%%%%%%%%%%%%%%%%%%%%%%%%%%%%%%%%%%%%%%%
%\newpage

%\begin{figure}[htbp]
%%\vspace{-0.8cm}
%\centerline{\psfig{figure=neoeventext3.ps,width=15cm}}
%%\vspace{-0.5cm}
%\caption{
%Same as Fig.12 but GRBR3 is used.
%}
%\label{event3}
%\end{figure}

%%%%%%%%%%%%%%%%%%%%%%%%%%%%%%%%%%%%%%%%%%%%%%%%%%%%%%%%%%%%%%%%%%%%%%
\newpage

\newpage

%%%%%%%%%%%%%%%%%%%%%%%%%%%%%%%%%%%%%%%%%%%%%%%%%%%%%%%%%%%%%%%%%%%%%%
%%%%%%%%%%%%%%%%%%%%%% TABLES %%%%%%%%%%%%%%%%%%%%%%%%%%%%%%%%%%%%%%%%
%%%%%%%%%%%%%%%%%%%%%%%%%%%%%%%%%%%%%%%%%%%%%%%%%%%%%%%%%%%%%%%%%%%%%%

\begin{table}
\caption{\label{tab:table0} Event Rate (yr$^{-1}$) and Statistical Error
of ANB at SK/TITAND}
\begin{tabular}{lcccllcccccccc}
    & $N$ (yr$^{-1}$) & $\sqrt{N \rm (yr^{-1}) }$\\ \hline
    $E_{\bar{\nu}_e}$ = (50~--~60)MeV & 1.0/1.0$\times 10^2$ &
    1.0/1.0$\times 10^1$\\ $E_{\bar{\nu}_e} \ge$ 50MeV &
    2.6/2.6$\times 10^2$ & 1.6/1.6$\times 10^1$\\ 
\end{tabular}
\end{table}

\begin{table}
\caption{\label{tab:table1} Event Rate (yr$^{-1}$) of Signals at SK/TITAND}
\begin{tabular}{ccccllcccccccc}
$\dot{M}$ ($M_{\odot}$s$^{-1}$) & $f$  & Oscillation & GRBR
& $E_{\bar{\nu}_e} = $(50~--~60)MeV 
& $E_{\bar{\nu}_e} \ge $50MeV \\
\hline
10 & 0.1 & Yes & 3 
& 9.1$\times 10^{-3}$/9.1$\times 10^{-1}$  
& 1.7$\times 10^{-2}$/1.7\\
10 & 0.5 & Yes & 3 
& 4.5$\times 10^{-2}$/4.5  
& 8.6$\times 10^{-2}$/8.6\\
10 & 1.0 & Yes & 3 
& 9.1$\times 10^{-2}$/9.1
& 1.7$\times 10^{-1}$/1.7$\times 10^{1}$\\
10 & 0.1 & No & 3 
& 1.8$\times 10^{-2}$/1.8
& 3.4$\times 10^{-2}$/3.4\\
10 & 0.5 & No & 3 
& 9.0$\times 10^{-2}$/9.0
& 1.7$\times 10^{-1}$/1.7$\times 10^{1}$\\
10 & 1.0 & No & 3 
& 1.8$\times 10^{-1}$/1.8$\times 10^{1}$  
& 3.4$\times 10^{-1}$/3.4$\times 10^{1}$\\
5 & 0.1 & Yes & 3 
& 4.6$\times 10^{-3}$/4.6$\times 10^{-1}$  
& 8.3$\times 10^{-3}$/8.3$\times 10^{-1}$\\
5 & 0.5 & Yes & 3 
& 2.2$\times 10^{-2}$/2.2
& 4.2$\times 10^{-2}$/4.2\\
5 & 1.0 & Yes & 3 
& 4.6$\times 10^{-2}$/4.6
& 8.3$\times 10^{-2}$/8.3\\
5 & 0.1 & No & 3 
& 9.1$\times 10^{-3}$/9.1$\times 10^{-1}$  
& 1.7$\times 10^{-2}$/1.7\\
5 & 0.5 & No & 3 
& 4.6$\times 10^{-2}$/4.6
& 8.3$\times 10^{-2}$/8.3\\
5 & 1.0 & No & 3 
& 9.1$\times 10^{-2}$/9.2
& 1.7$\times 10^{-1}$/1.7$\times 10^{1}$\\
1 & 0.1 & Yes & 3 
& 8.3$\times 10^{-4}$/8.3$\times 10^{-2}$  
& 1.4$\times 10^{-3}$/1.4$\times 10^{-1}$\\
1 & 0.5 & Yes & 3 
& 4.2$\times 10^{-3}$/4.2$\times 10^{-1}$  
& 6.9$\times 10^{-3}$/6.9$\times 10^{-1}$\\
1 & 1.0 & Yes & 3 
& 8.3$\times 10^{-3}$/8.3$\times 10^{-1}$  
& 1.4$\times 10^{-2}$/1.4\\
1 & 0.1 & No & 3 
& 1.7$\times 10^{-3}$/1.7$\times 10^{-1}$  
& 2.7$\times 10^{-3}$/2.7$\times 10^{-1}$\\
1 & 0.5 & No & 3 
& 8.3$\times 10^{-3}$/8.3$\times 10^{-1}$  
& 1.4$\times 10^{-2}$/1.4\\
1 & 1.0 & No & 3 
& 1.7$\times 10^{-2}$/1.7
& 2.7$\times 10^{-2}$/2.7\\
\end{tabular}
\end{table}

\begin{table}
\caption{\label{tab:table2} Time (yr) Required to Detect the Signals
at SK/TITAND}
\begin{tabular}{ccccllcccccccc}
$\dot{M}$ ($M_{\odot}$s$^{-1}$) & $f$  & Oscillation & GRBR
& $E_{\bar{\nu}_e} = $(50~--~60)MeV  
& $E_{\bar{\nu}_e} \ge $50MeV \\
\hline
10 & 0.1 & Yes & 3 
& 1.3$\times 10^{4}$/1.3$\times 10^{2}$  
& 2.3$\times 10^{4}$/2.3$\times 10^{2}$\\
10 & 0.5 & Yes & 3 
& 5.1$\times 10^{2}$/5.1  
& 9.3$\times 10^{2}$/9.3\\
10 & 1.0 & Yes & 3 
& 1.3$\times 10^{2}$/1.3
& 2.3$\times 10^{2}$/2.3\\
10 & 0.1 & No & 3 
& 3.1$\times 10^{3}$/3.1$\times 10^{1}$  
& 5.8$\times 10^{3}$/5.8$\times 10^{1}$\\
10 & 0.5 & No & 3 
& 1.3$\times 10^{2}$/1.3  
& 2.3$\times 10^{2}$/2.3\\
10 & 1.0 & No & 3 
& 3.2$\times 10^{1}$/3.2$\times 10^{-1}$  
& 5.8$\times 10^{1}$/5.8$\times 10^{-1}$\\
5 & 0.1 & Yes & 3 
& 5.0$\times 10^{4}$/5.0$\times 10^{2}$  
& 9.9$\times 10^{4}$/9.9$\times 10^{2}$\\
5 & 0.5 & Yes & 3 
& 2.0$\times 10^{3}$/2.0$\times 10^{1}$  
& 4.0$\times 10^{3}$/4.0$\times 10^{1}$\\
5 & 1.0 & Yes & 3 
& 5.0$\times 10^{2}$/5.0
& 9.9$\times 10^{2}$/9.9\\
5 & 0.1 & No & 3 
& 1.3$\times 10^{4}$/1.3$\times 10^{2}$  
& 2.5$\times 10^{4}$/2.5$\times 10^{2}$\\
5 & 0.5 & No & 3 
& 5.0$\times 10^{2}$/5.0  
& 9.9$\times 10^{2}$/9.9\\
5 & 1.0 & No & 3 
& 1.3$\times 10^{2}$/1.3  
& 2.5$\times 10^{2}$/2.5\\
1 & 0.1 & Yes & 3 
& 1.5$\times 10^{6}$/1.5$\times 10^{4}$  
& 3.7$\times 10^{6}$/3.7$\times 10^{4}$\\
1 & 0.5 & Yes & 3 
& 6.0$\times 10^{4}$/6.0$\times 10^{2}$  
& 1.5$\times 10^{5}$/1.5$\times 10^{3}$\\
1 & 1.0 & Yes & 3 
& 1.5$\times 10^{4}$/1.5$\times 10^{2}$  
& 3.7$\times 10^{4}$/3.7$\times 10^{2}$\\
1 & 0.1 & No & 3 
& 3.8$\times 10^{5}$/3.8$\times 10^{3}$  
& 9.2$\times 10^{5}$/9.2$\times 10^{3}$\\
1 & 0.5 & No & 3 
& 1.5$\times 10^{4}$/1.5$\times 10^{2}$  
& 3.7$\times 10^{4}$/3.7$\times 10^{2}$\\
1 & 1.0 & No & 3 
& 3.8$\times 10^{3}$/3.8$\times 10^{1}$  
& 9.2$\times 10^{3}$/9.1$\times 10^{1}$\\
\end{tabular}
\end{table}

\end{document}